\documentclass[
  journal=pasa,
  manuscript=research-paper, 
  year=2023,
  volume=XX,
]{cup-journal}

\usepackage{microtype,siunitx,booktabs,rotating,soul,hyperref}
\sethlcolor{yellow}

\title{Survey-scale discovery-based research processes: evaluating a bespoke visualisation environment for astronomical survey data}

\author{C. J. Fluke}
\affiliation{Centre for Astrophysics \& Supercomputing, 
Swinburne University of Technology, Hawthorn, 3122, Australia}
\email[C. J. Fluke]{cfluke@swin.edu.au}

\author{D. Vohl}
\affiliation{Anton Pannekoek Institute, University of Amsterdam, Postbus 94249, 1090 GE Amsterdam, The Netherlands}
\alsoaffiliation{ASTRON, Netherlands Institute for Radio Astronomy, Oude Hoogeveensedijk 4, 7991 PD, Dwingeloo, The Netherlands}

\author{V. A. Kilborn}
\affiliation{Centre for Astrophysics \& Supercomputing, 
Swinburne University of Technology, Hawthorn, 3122, Australia}

\author{C. Murugeshan}
\affiliation{CSIRO, Space and Astronomy, PO Box 1130, Bentley, WA 6102, Australia}
\alsoaffiliation{ARC Centre of Excellence for All Sky Astrophysics in 3 Dimensions (ASTRO 3D), Australia}

\doi{XXX}

\received {dd Mmm YYYY}
\revised  {dd Mmm YYYY}
\accepted {dd Mmm YYYY}
\published{dd Mmm YYYY}

\keywords{data-intensive astronomy -- visual discovery} 

\begin{document}

\begin{abstract}
Next generation astronomical surveys naturally pose challenges for human-centred visualisation and analysis workflows that currently rely on the use of standard desktop display environments.  While a significant fraction of the data preparation and analysis will be taken care of by automated pipelines, crucial steps of knowledge discovery can still only be achieved through various level of human interpretation.   As the number of sources in a survey grows, there is need to both modify and simplify repetitive visualisation processes that need to be completed for each source.  As tasks such as per-source quality control, candidate rejection, and morphological classification all share a single instruction, multiple data (SIMD) work pattern, they are amenable to a parallel solution.  Selecting extragalactic neutral hydrogen (H{\sc i}) surveys as a representative example, we use system performance benchmarking and the visual data and reasoning (VDAR) methodology from the field of information visualisation to evaluate a bespoke comparative visualisation environment: the {\tt encube} visual analytics framework deployed on the 83 Megapixel Swinburne Discovery Wall.   Through benchmarking  using spectral cube data from existing H{\sc i} surveys, we are able to perform interactive comparative visualisation via texture-based volume rendering of 180 three-dimensional (3D) data cubes at a time. The time to load a configuration of spectral cubes scale linearly with the number of voxels, with independent samples of 180 cubes (8.4 Gigavoxels or 34 Gigabytes) each loading in under 5 minutes.  We show that parallel comparative inspection is a productive and time-saving technique which can reduce the time taken to complete SIMD-style visual tasks currently performed at the desktop by at least two orders of magnitude,  potentially rendering some labour-intensive desktop-based workflows obsolete.
\end{abstract}

\section{Introduction}
Next generation astronomical surveys will pose challenges for a range of human-centred visualisation and analysis workflows that currently rely on the use of standard desktop display environments. Knowledge discovery activities that were, or perhaps still are, feasible for a human to perform when the quantity (i.e. volume) or rate (i.e. velocity) of data available was low are becoming more reliant on automated or autonomous solutions.  

While desktop computing has already been augmented through the adoption of supercomputing and cloud-style remote services, the visualisation and display of astronomical data is still strongly dependent on the utilisation of laptop screens or monitors located in the astronomer's office.

To address the specific needs of individual astronomers, and astronomical research teams, a collection of data analysis and visualisation tools are required. This includes continuing to take full advantage of existing, well-established options that are able to be scaled-up effectively, along with developing and assessing the potential of novel solutions or systems that either provide extra functionalities, or that can be connected into extensible workflows (e.g. virtual observatory model).

\subsection{Comparative visualisation}
Seeing many sources together -- comparative visualisation -- is an approach that naturally supports pattern-finding (``those galaxies all show similar kinematic properties'') and anomaly detection (``why is that one source so different to everything else?'').

Such multi-object comparisons might include quality control activities (e.g. assessing whether a source finder or automated calibration pipeline is functioning as expected by selecting a sample of sources for assessment, which might include fine-tuning to check or verify a machine learning algorithm), investigating outcomes of model-fitting (e.g. examining the residual signal once different types of kinematic models are applied), or any of a range of standard analysis tasks that can be performed based on morphological or environmental selection criteria (e.g. field compared with cluster galaxies, dwarf galaxies versus grand design spirals, or the discovery of novel classes of objects when a new discovery space is opened).  We will refer to all such activities as {\em survey-scale discovery-based research processes}, as the purpose is to explore data in order to make sense of it [see the model of ``sensemaking'' presented by \citet{Pirolli05}, and applied in Section \ref{sct:vdar}].

\begin{figure*}[ht]
	\includegraphics[width=1.0\columnwidth]{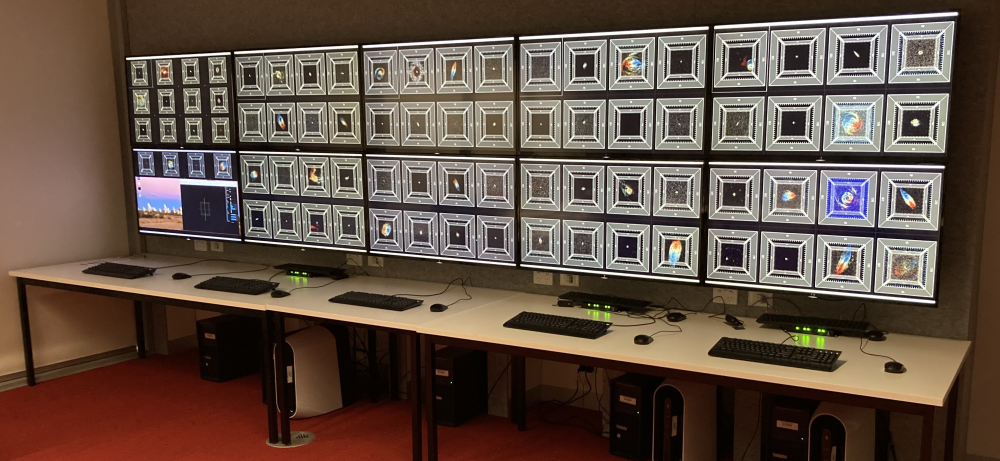}
    \caption{The Swinburne Discovery Wall: a multi-purpose 83 Megapixel tiled display wall, comprising a matrix of two rows and five columns of Philips BDM4350UC 4K-UHD monitors and five Lenovo ThinkStation P410 MiniTowers.   See Section \ref{sct:dwall} and Table \ref{tbl:techdef} for additional details.   A small-multiples visualisation approach is used, with a single-instruction multiple data interaction paradigm. Interaction with the dataset is achieved through the browser-based user interface, visible in the left-hand monitor in the bottom row.
Columns are enumerated from 1 to 5 from left to right.  The keyboards in front of each column can be used for direct interaction with an individual data cube on the corresponding column.  Shown here is a configuration of 80 spectral cubes sampled from the WHISP \citep{Hulst2001ASPC..240..451V,Swaters02}, THINGS \citep{Walter2008AJ} and LVHIS \citep{Koribalski2018MNRAS.478.1611K} projects (see Section \ref{sct:surveydata}). }
    \label{fig:sdw}
\end{figure*}

Limited scope for comparative visualisation can occur by either loading data into several independent instances of a visualisation tool (usually on the same computing platform) or by switching between individual views of multiple objects, requiring loading and unloading of data.
 When working with large-scale survey data, desktop-based visualisation strategies may lead to a reduction in the ability for an individual to see patterns across a sizeable portion of the survey.

In practice, effective comparative visualisation cannot be achieved by moving between visualisations of one or two objects at a time. At each stage, there is a loss of time to input/output, and a strong reliance on the visual recall abilities of the astronomer [see \citet{Norris94} for a related discussion].  Individual instances are unlikely to have linked camera actions (e.g. panning, rotation, zoom, scaling), requiring the use of repetitive interaction processes.
Moreover, if performed at the desktop, the small physical display space of a standard monitor is not always conducive to real-time, collaborative inspection for those researchers who prefer, or find it more productive, to work this way.

\subsection{Single instruction, multiple data work patterns}
Survey-scale discovery-based research processes, such as those described above, are all highly repetitive, and may need to be completed for each individual source.  Many repetitive research processes share a single instruction, multiple data (SIMD) work pattern, and so are amenable to a parallel solution.

One approach to the parallelisation of human-centred visualisation and analysis tasks is to share the work out amongst multiple team members [e.g. as occurred while preparing catalogues for the H{\sc i} Parkes All Sky Survey -- see \citet{Barnes01} and \citet{Meyer04}], or further afield via crowd-sourcing of citizen scientists \citep[e.g.][]{Banfield+2015}.  

A limitation to these distributed processes is one of consistency in decision-making between team members with diverse skill levels [see, for example, \citet{Fluke17,Fluke2020}].  An investment in training may be required,  or a complex task must be abstracted to one of group-consensus classification.  Furthermore, while serendipitous discoveries do occur in citizen science activities, that is not the norm.  

An alternative is to change the viewing paradigm, so that a more suitable mode of parallel inspection by a single researcher, or co-located team, can be achieved.  This is the approach we investigate in this work using {\tt encube}\footnote{Long term access to open source software described by \citet{encube2017ascl.soft06007V}.}: a visual analytics framework for collaborative and comparative visualisation, designed to work on a multi-monitor tiled display wall and dedicated compute nodes \citep{Vohl16}. Figure \ref{fig:sdw} shows {\tt encube} operating on the Swinburne Discovery Wall (see Section \ref{sct:dwall}), providing simultaneous display of 80 spectral cubes sampled from three extragalactic neutral hydrogen (H{\sc i}) surveys (described in more detail in Section \ref{sct:surveydata}).

\subsection{The visual data analysis and reasoning methodology}
In order to best utilise non-standard or novel visualisation systems, it is important to understand their strengths and weaknesses.  The suitability of any visualisation approach or environment -- software or hardware, standard or bespoke -- should be examined or evaluated using appropriate methodologies. 

Looking to the broader field of information visualisation, such evaluations can include investigation of either the {\em process} of visualisation or the {\em nature} of visualisation systems and algorithms \citep{Lam12,Isenberg13}.  For our investigation of survey-scale discovery-based research processes, we select the empirical visual data analysis and reasoning (VDAR) methodology.

A VDAR evaluation is usually approached via a case study: a cohort of experts assess {\em their} ability to derive knowledge about a relevant dataset while using a new visualisation system, software or strategy to  perform domain-specific tasks \citep{Lam12,Isenberg13}.  

As our relevant dataset, we utilise existing extragalactic H{\sc i} survey data (see Section \ref{sct:surveydata}), available as an ensemble of spectral cubes (two spatial dimensions and one spectral dimension). We consider three representative survey-scale discovery-based research processes that can occur in the preparation and analysis of large-scale extragalactic H{\sc i} surveys:
\begin{enumerate}
    \item  {\em Quality control} of individual sources, ensuring that calibrations have been applied correctly and bad channels (e.g. impacted by interference or instrumental features) have been flagged or removed;
    \item {\em Candidate rejection}, whereby false-positive detections from automated source finders are identified and removed from the catalogue. This can also help to improve training sets of ``non-source'' examples for use with machine learning and related automated methods; and 
    \item {\em Morphological classification}, identifying and sorting sources into categories based on observed structural, kinematic or environmental properties.  The classification process may also include anomaly detection, wherein unexpected discoveries are made based on the observed structural properties.
\end{enumerate}

Through a mix of visual analytic functionalities, including interactive three-dimensional (3D) volume rendering methods, {\tt encube} provides ways to explore both spatial and spectral features, which can be matched to other observed or derived parameters. A 3D approach can help to reveal complex kinematic structures or system artefacts that might otherwise appear only in projection using moment maps or position-velocity diagrams.

We choose to perform our evaluation with 3D methods as they: (1) are the current defaults within the public {\tt encube} code; (2) present an upper bound in terms of the computation required for benchmarking purposes; and (3) provide the VDAR user cohort with access to novel comparative sensemaking strategies via the Swinburne Discovery Wall.  For other applications, alternative data visualisation modes such as moment maps\footnote{A camera projection parallel to any axis of a spectral cube can be used to generate a two-dimensional (2D) projection of the data~\citep[][Figure A1]{vohl2017MNRAS.471.3323V}, and hence can be used to generate 2D solution space representations while still retaining access to the full representation of the data in memory for fast calculations using graphics shaders.}  or scatter plots could be utilised as they are supported by the underlying visualisation framework.

\subsection{Overview}
   In this paper, we consider a specific visualisation problem that is not feasible to address using a desktop-based visualisation solution: interactive, comparative visualisation of ${\geq\!100}$ data instances.  We evaluate the practicality of using a bespoke visualisation environment (viz.  {\tt encube} and the Swinburne Discovery Wall) for survey-scale discovery-based research processes through: (1) system benchmarking, which provides quantitative information on system performance and scalability; and (2) a visual data analysis and reasoning study.

For five different display configurations, supporting simultaneous visualisation of 20, 40, 80, 120 or 180 spectral cubes, selected from representative extragalactic H{\sc i} survey datasets, we report benchmarking in terms of the two most critical factors: (1) the time taken to load an ensemble of spectral cubes; and (2) the typical minimum interactive frame rate.  
Together, these values allow us to estimate the visualisation throughput, $V_{\rm tp}$ (sources/hour), that might be achieved by a single user when undertaking SIMD tasks such as quality control, candidate rejection or morphological classification.

Compared to the serial case of viewing one data instance at a time on a standard desktop monitor, {\tt encube} and the Swinburne Discovery Wall could decrease the time taken to complete survey-scale comparative visualisation workflows by a factor of 100 or more.

In Section \ref{sct:technical}, we explain the main technical elements of the bespoke visualisation environment. In Section \ref{sct:casestudy}, we provide background on the extragalactic H{\sc i} case study.   
We evaluate the visualisation environment through system benchmarking (Section \ref{sct:benchmarks}) and via the VDAR evaluation (Section \ref{sct:vdar}), which considers three typical discovery-based SIMD activities: quality control, candidate rejection and morphological classification.   
We present a discussion of our finding in Section \ref{sct:discuss}, and present our conclusions in Section \ref{sct:conclude}.
Further technical and implementation notes can be found in  \ref{app:technical}.

Our approach can be generalised to any survey datasets comprising more individual observations or instances than can be comfortably analysed or scrutinised by one investigator on a standard desktop display. This might include two-dimensional images or moment-map  projections, optical/infrared spectral cubes (e.g. from integral field spectroscopy), or simulation data products. The comparative visualisation strategies demonstrated here are applicable to any similar SIMD-style activity, and are not restricted to the specific use of {\tt encube} with the Swinburne Discovery Wall.  As an open source solution, users are encouraged to modify the functionality of {\tt encube} (e.g. in order to provide alternative 2D or 3D visualisation modes or to handle domain-specific data formats) or reconfigure the arrangement of the display environment to suit their own survey-scale discovery-based research needs.

\section{A bespoke comparative visualisation environment}
\label{sct:technical}
In this section, we provide a technical overview of the two main components of the bespoke comparative visualisation environment used in this work: (1) the {\tt encube} framework, which enables visualisation of multiple data instance (in the form of spectral cubes for our case study); and (2) the Swinburne Discovery Wall, a specific instance of a large-area tiled display wall.

{\tt Encube} was conceptualised and developed specifically to support SIMD visualisation and analysis tasks, with an aim to accelerate data-intensive comparative visualisation and discovery workflows.  {\tt Encube} displays multiple individual data visualisations across single or multiple display devices, with interaction coordinated through a user interface on the master node.   For related approaches, see the virtual reality implementation of {\tt BentoBox} \citep[][and references therein]{Johnson19} and the ``shelves'' metaphor for small-multiples that considers utilisation of immersive space \citep{Liu20}.

\subsection{The {\tt encube} framework}
The {\tt encube} framework \citep{Vohl16,Vohl2017IAUS..325..311V} supports comparative visualisation and analysis of survey data (also referred to as an ensemble in other domains). The primary development emphasis was for structured 3D data: spectral cube data from astronomy and magnetic resonance imaging data from medical imaging. {\tt Encube} provides an interactive data exploration and analysis experience, employing a strategic mixture of software (data processing, management, visualisation, analysis) and hardware (graphics processing units, computer cluster, displays). 

{\tt Encube} is a modular and open-source code base \citep{encube2017ascl.soft06007V}, where each module targets a specific set of tasks within a visual analytics workflow: (1) processing and visualisation of data; (2) workflow and communication management; and (3) user interactions. Similar to a microservices-style architecture, the modular design allows individual components to be connected, enhanced or replaced as required, so that {\tt encube} can be kept compatible with, and scalable to, the requirements of future science operations. For instance, customisable code for 3D visualisation is currently created using the C/C++ languages for good performance with the {\tt S2PLOT} interactive programming library \citep{Barnes2006PASA...23...82B}, which builds on the OpenGL\footnote{\url{http://www.opengl.org}} graphics library.

From a system architecture standpoint, {\tt encube} comprises a process layer and an input/output (I/O) layer.  The process layer performs data processing tasks (load data, compute statistics, render visualisation), and the I/O layer responds to user inputs and generates visual outputs.  Each layer contains units where specified tasks are performed.  Depending on the task, a unit can be instantiated once, or multiple times for parallel operation (generally on different compute hardware).  In its current form, the {\tt encube} process layer comprises a single {\em manager} unit and one or more {\em process and render} units, while the I/O layer contains an interaction unit and one or more display units. 

Units can communicate between each other in order to pass workflow information across the architecture.  The communication pathway between units can be represented as a directed graph [see Figures 2 and 4 of \citet{Vohl16}]: 

\begin{eqnarray}
&\mbox{Interaction unit(s)} \nonumber \\
&\updownarrow \nonumber \\
&\mbox{Manager unit} \nonumber \\
&\updownarrow \nonumber \\
&\mbox{Process and Render unit(s)} \nonumber \\
&\downarrow \nonumber \\
&\mbox{Display unit(s)} \nonumber
\end{eqnarray}
where the arrows indicate the information flow direction between two unit vertices on the graph. Based on the number of instances of a unit, communication can include serial or parallel messages. We note that peer-to-peer communication within a unit type is not currently implemented (e.g. direct message passing between two interaction units).

The manager unit orchestrates the overall software workflow.  It first reads a configuration file containing network information about the available compute nodes, characteristics of the tiled visualisation output, along with system metadata and the location of the dataset. This unit also schedules and synchronises the workflow, sharing metadata as well as commands with other neighbouring units. Here, the manager unit acts as a messenger between an interaction unit and a process and render unit. Moreover, given that all commands pass through the manager unit, the workflow history and system state can be recorded (if requested) so that actions can be revised, replicated, or continued later. 

The interaction unit is where a user interacts with the dataset. In particular, the user can specify which data files to load and visualise, change visualisation parameters (e.g. ray-tracing method), select and organise individual visualisations, and request diagnostic plots.   The interaction unit provides a ``world in miniature'' view of the display setup, mapping regions within the user interface to the physical display. 

Metadata is presented in a table, which can be sorted by categories. Visualisations are generated after selecting rows of the table, either individually or by ordered batch (e.g. sorted by parameters such as distance, size, etc.).  Once data is loaded into memory on a process and render unit, visualisation parameters (e.g. histogram thresholds, spatial cropping, colourmap selection) can be updated in real time to modify one or more visualisations. Global or partial statistical values can also be computed on request for selected data files and gathered to summarise properties of a subset. 

The process and render unit provides functionalities such as loading data files to GPU memory, computing statistics (e.g. mean, standard deviation, histogram), creating  visualisation callbacks (e.g. including responses to input via keyboard, mouse, or the remote user interface), and generating the visualisations through texture-based volume rendering. 

Finally, a visualisation rendered by a process and render unit is displayed on screen via the display unit. A display unit provides a mapping to one or more physical screens via the configuration file read by the manager unit.

\subsection{The Swinburne Discovery Wall}
\label{sct:dwall}
From its inception, {\tt encube} was designed for use in high-end visualisation environments comprising multiple off-the-shelf displays, i.e. a tiled display wall (TDW). See \citet{Meade14} and \cite{Pietriga16} for detailed investigations of the role of TDWs in astronomy. A TDW provides several advantages over a standalone workstation monitor: many more pixels, a greater display area, and, in some cases, access to additional co-located computing power.

Initial deployment and testing of {\tt encube} was undertaken with the CAVE2$^{\rm TM}$ hybrid high-performance computing and visualisation space at Monash University [as reported in \citet{Vohl16}].  The Monash CAVE2$^{\rm TM}$ \citep{Sommer17} comprised 80 stereoscopic-capable displays, with a cylindrical configuration (330 degrees to allow entry and exit from the physical space) of four rows and 20 columns.  Collectively, the environment provided 84 million pixels for two-dimensional display and 42 million pixels in stereoscopic mode.  The Monash CAVE2$^{\rm TM}$ was linked to a real-time compute cluster with a peak of 100 Tflop/s and 240 GB of GPU memory.

Additional development, and the activities presented in this work, utilised the Discovery Wall (Figure \ref{fig:sdw}) operated at Swinburne University of Technology.  The Swinburne Discovery Wall is a TDW comprising ten Philips BDM4350UC 4K ultra high-definition (4K-UHD) monitors
arranged in a matrix of two rows and five columns.  The total pixel count is approximately 83 Megapixels and the accessible screen area is just under 5.0 m$^{2}$ (see Table \ref{tbl:techdef}).

\begin{table}[ht]
	\centering
	\caption{Specifications for the ten Philips BDM4350UC 4K-UHD monitors of the Swinburne Discovery Wall. Parameters and corresponding units are: screen linear dimension, $L_{\rm dim}$ (m $\times$  m), screen area, $A_{\rm screen}$ (m$^2$), pixel dimensions, $P_{\rm dim}$ (pix $\times$ pix), and total pixels, $P_{\rm total}$ (Megapixels).  
	}
\label{tbl:techdef}

\begin{tabular}{lcccc}
& 
$\mathbf{L_{\rm dim}}$     &  
$\mathbf{A_{\rm screen}}$     & 
$\mathbf{P_{\rm dim}}$     & 
$\mathbf{P_{\rm total}}$ \\\hline
Per screen    & $0.94 \times 0.53$ & $0.499$ & $3840 \times 2160$ & $8.29$ \\ 
Per column    &  $0.94 \times 1.06$ & $0.996$ & $3840 \times 4320$ & $16.59$\\
Discovery Wall & $4.70 \times 1.06$ & $4.982$ & $19200 \times 4320$ & $82.94$  \\
\end{tabular}

\end{table}

Each column of the Discovery Wall is connected to a Lenovo ThinkStation P410 Mini Tower (2.8 GHz, 16 GB RAM) with an NVIDIA GTX1080 graphics card (8 GB). The workstations operate with the CentOS\footnote{\url{http://www.centos.org}} Linux operating system (Version 7.4.1708), noting that we use the version of CentOS that was installed on the Discovery Wall when it was commissioned in 2018.

The original iteration of the Swinburne Discovery Wall, which operated until November 2021, had one additional column of two 4K-UHD monitors such that the total screen area was 6.0 m$^2$ and a pixel count closer to 1 million pixels.  In December 2021, the Discovery Wall hardware was transferred to a new location, but with insufficient wall-space to accommodate all six columns. Reconfiguration of {\tt encube} to work on the relocated and reduced-scale Discovery Wall in February 2022 required approximately two minutes to remove references to the sixth Lenovo MiniTower workstation from the {\tt encube} source and scripts.

\section{Case study: extragalactic H{\sc i} atronomy}
\label{sct:casestudy}
Consider the specific case of extragalactic H{\sc i} astronomy, which is based on observations of the 21 cm (1420.40576 MHz) hyperfine spin flip transition of the hydrogen atom. Theoretically predicted by \citet{vandeHulst1945}, and first detected by \citet{EwenPurcell1951}, \citet{MullerOort1951} and \citet{Pawsey1951}, the 21 cm line provides a valuable signature of the neutral gas content of galaxies. 

Apart from being the primary component from which stars are eventually formed, the H{\sc i} gas in galaxies is also typically much more extended than their stellar discs [see \citet{Verheijen01}] making it an important tracer of the effects of both internal properties of galaxies, such as feedback and angular momentum \citep{Genel15,Obreschkow16,Murugeshan20}, as well as environmental processes such as ram pressure and tidal stripping to name a few [see \citet{Gunn72} and \citet{Fasano00}]. For these reasons, high spatial and spectral resolution studies of the HI gas distribution in galaxies are paramount for our understanding of galaxy evolution.

Historically, extragalactic H{\sc i} surveys fall into three broad categories: (1) spectral line observations, using single-dish radio telescopes; (2) spatial mapping with multi-beam receivers \citep[e.g.][]{StaveleySmith96}, whereby it became feasible to undertake spectral-line surveys at a large scale \citep{Barnes01};
 and (3) high-resolution spectral cube observations, utilising aperture synthesis.  
 
\subsection{Extragalactic neutral hydrogen surveys}
The number of sources available from H{\sc i} surveys is undergoing a step-change. New wide-field and deep surveys have been enabled through instruments and facilities including:
\begin{itemize}
    \item The APERture Tile In Focus (APERTIF) upgrade to the Westerbork Synthesis Radio Telescope (WRST) -- see \citet{Verheijen08}, with H{\sc i} survey descriptions in \citet{Verheijen2009pra..confE..10V}, \citet{AdamsvanLeeuwen2019} and \citet{Adams+2022}; 
    \item The Australian Square Kilometre Array Pathfinder (ASKAP) -- see \citet{Johnston07,Johnston08,Hotan2021} and H{\sc i} survey descriptions for the Widefield ASKAP L-band Legacy All-sky Blind SurveY (WALLABY) in \citet{KoribalskiStaveleySmith2009WALLABY} and \citet{Koribalski20}; and
    \item MeerKAT \citep{Booth09}, with local \citep{deBlok17} and ultra-deep \citep{Holwerda2012IAUS..284..496H} H{\sc i} surveys planned.
\end{itemize}
The scale and rate of data collection from these programs provide a first opportunity to prepare for the future of H{\sc i} astronomy that will occur with the Square Kilometer Array (SKA).

Using WALLABY as an example, these surveys will produce three main categories of data: 
\begin{enumerate}
    \item Large-scale {\em survey cubes}. Over a period of five years, WALLABY is expected to cover up to $1.4\pi$ sr of the sky with $\sim 550$ full-resolution spectral cubes.   Each cube is anticipated to have $4200 \times 4200$ spatial pixels and $7776$ spectral channels, requiring $\sim 600$ Gigtabytes (GB) per cube.  The total data storage required for WALLABY will exceed 1 Petabyte. 
    \item Small-scale {\em source cubelets}.  By running the Source Finding Application \citep[SoFiA;][]{Serra15,Westmeier21} on the survey cubes, candidate source cubelets  can be extracted and stored separately, or simply have the coordinates of their bounding boxes within the survey cubes stored [see \citet{Koribalski2012PASA} for an overview, and \citet{Popping12} for a comparison of H{\sc i} source finders]. As source cubelets take up only a small fraction of the survey cubes, this is a much more manageable data volume to work with.  Estimates of the number of H{\sc i} detections from WALLABY exceed 200,000 sources. Approximately $15$--$20 \%$ of these sources are expected to be spatially resolved (i.e. where the spatial distribution of H{\sc i} is visible, which is anticipated to require at least 3-4 resolution elements or synthesised beams across the source).
    \item Catalogues of {\em derived data products}. Along with the key parameters (e.g. position, velocity dispersion, H{\sc i} flux) generated by source finders such as SoFiA and Selavy \citep{Whiting2012},   further automated processing and analysis tasks can provide additional data. This includes activities such as disk-based model fitting [e.g. {\tt TiRiFiC} \citep{Jozsa07}, $^{\rm 3D}${\tt BAROLO} \citep{DiTeodoro15}, or {\tt 2DBAT}, \citep{Oh18}, and see also the description of the WALLABY Kinematic Analysis Proto-Pipeline (WKAPP) in \cite{Deg22}], computation of integral properties (e.g. total H{\sc i} mass, star formation rates), or cross-matching with optical/infrared catalogues. 
\end{enumerate}
Each of these data products will aid  the development of insight and improved understanding of H{\sc i}'s role in galaxy formation and evolution.  

\subsection{Visualisation-dominated workflows}
\label{sct:visdom}
The data-intensive demands of new H{\sc i} surveys has motivated the development of a number of customised tools for interactive qualitative and quantitative spectral cube visualisation \citep{Hassan11,Lan21}.  

Moving beyond the well-established and widely-utilised solutions such as {\tt Karma}\footnote{https://www.atnf.csiro.au/computing/software/karma} \citep{Gooch96} and {\tt CASA}\footnote{https://casa.nrao.edu} [the Common Astronomy Software Applications package; \citet{McMullin07}], 
alternatives for desktop-based visualisation and analysis include {\tt AstroVis} \citep{Perkins14}, {\tt SlicerAstro} \citep{Punzo15,Punzo16,Punzo17}, {\tt FRELLED} [\citet{Taylor15} using the free, open-source {\tt Blender} animation software], {\tt FITS3D} \citep{Mohan17}, {\tt Shwirl} \citep{vohl2017MNRAS.471.3323V}, and {\tt CARTA}\footnote{https://cartavis.org/} \citep[Cube Analysis and Rendering Tool for Astronomy;][]{Comrie21}.

\citet{Ferrand16} prototyped a solution using the Unity\footnote{https://unity.com} real-time 3D engine, which can be deployed on a desktop or operate with a variety of advanced display technologies.   With their {\tt iDAVIE} solution, \citet{Jarrett21} have successfully moved spectral cube visualisation and analysis into interactive and immersive virtual reality environments.  

Finally, targeting data products that greatly exceed the processing capabilities of standard desktop computers, \citet{Hassan-2013} achieved real-time interactive visualisation of Terabyte-scale spectral cubes using a high-performance solution with graphics processing units (GPUs) and the {\tt GraphTIVA} framework.

For most of these examples, the workflow for visualisation and analysis of the gas in galaxies emphasises the study of one galaxy at a time.  When the data volume is low and the data rate is slow, a great deal of human time can be dedicated to examining individual data cubes or source cubelets.   While highly appropriate in an era of small surveys, this serial processing presents a bottleneck for knowledge discovery once the ASKAP and MeerKAT surveys scale up to include many thousands of spatially resolved sources.

The transformation of a survey cube to a subset of source cubelets, and ultimately, a reliable, science-ready catalogue of data products can be encapsulated as a workflow.  Parts of the workflow are expected to be fully automated [e.g. the Apercal calibration pipeline for Apertif surveys \citep{Adebhar22} or ASKAPSoft for ASKAP \citep{Guzman19,Wieringa20}].  Other stages will rely on some level of human intervention, either through computational steering (selecting parameters for the workflow, setting thresholds on source finders, etc.) or data visualisation for analysis and discovery.

\begin{table*}[ht]
\begin{center}
\caption{Extragalactic H{\sc i} surveys used for evaluating {\tt encube} on the Swinburne Discovery Wall. $N_{\rm s}$ is the number of spectral cubes selected from each of the three surveys (see Section \ref{sct:surveydata} for a discussion as to why several spectral cubes were omitted). Data volumes are reported in Megabytes (MB) and voxel counts in Megavoxels (Mvox), with spectral cubes stored in the FITS format. Statistical quantities presented are the min(imum), max(imum), mean, sample standard deviation (SD) and median.  The total column summarises the volume or voxel count for the entire survey. } 

\label{tbl::surveys}
\begin{tabular}{cccrrcrr}
{\bf Parameter} & {\bf Survey} & $\mathbf{N_{\rm s}}$ & {\bf Min} & {\bf Max} & {\bf Mean} $\mathbf{\pm}$  {\bf SD} & {\bf Median} & {\bf Total} \\ \hline
Volume	&	WHISP	&	254	&	66.1	&	133.2	&	120.8 $\pm$ 25.8	&	133.2	&	30680	\\
(MB) &	THINGS	&	32	&	145.2	&	1169.8	&	378.7  $\pm$ 221.9	&	320.9	&	12119\\
		&	LVHIS	&	80	&	10.8	&	720.0	&	79.1  $\pm$ 91.6	&	57.6	&	6328	\\
\hline															
Voxels	&	WHISP	&	254	&	16.5	&	33.3	&	30.2  $\pm$ 6.5	&	33.3	&	7669	\\
(Mvox)	&	THINGS	&	32	&	35.7	&	289.4	&	92.5  $\pm$ 54.8	&	78.6	&	2960	\\
	&	LVHIS	&	80	&	2.7	&	180.0	&	19.8  $\pm$ 22.9	&	14.4	&	1582	\\
	\hline
\end{tabular}
\end{center}
\end{table*}

\subsection{Survey data}
\label{sct:surveydata}
While future applications of the comparative visualisation strategies examined here may include the H{\sc i} surveys to be conducted with ASKAP and MeerKAT, we perform the benchmarking and VDAR evaluations using data from three extant H{\sc i} surveys that targetted nearby spiral and irregular galaxies: 
\begin{enumerate}
\item WHISP: Westerbork Observations of Neutral Hydrogen in Irregular and Spiral Galaxies\footnote{http://wow.astron.nl}, undertaken with the Westerbork Synthesis Radio Telescope \citep{Hulst2001ASPC..240..451V,Swaters02};
\item THINGS: The H{\sc i} Nearby Galaxy Survey\footnote{https://www2.mpia-hd.mpg.de/THINGS/Data.html} comprising high-spectral and high-spatial resolution data from the National Radio Astronomy Observatory Very Large Array \citep{Walter2008AJ}; and
\item LVHIS: The Local Volume H{\sc i} Survey\footnote{https://www.atnf.csiro.au/research/LVHIS/LVHIS-database.html}, which obtained deep H{\sc i} line and 20-cm radio continuum observations with the Australia Telescope Compact Array  \citep{Koribalski2018MNRAS.478.1611K}.
\end{enumerate}

\begin{table*}[ht]
\caption{Display and survey configurations for which the {\tt encube} benchmarks were obtained. Set is the label used to identify the five different configurations (A-E), with $N_{\rm cube}$ = 20, 40, 80, 120 or 180.  Config is the arrangement of {\tt S2PLOT} panels (rows $\times$ columns) per column of the Discovery Wall.     Survey is one of [W]HISP, [T]HINGS, [L]VHIS or [C]ombination. $N_{\rm W}$, $N_{\rm T}$, and $N_{\rm L}$ are the number of spectral cubes selected from each of the input surveys.  Random sampling with replacement is used for configurations where the total number of cubes displayed exceeds the input survey size. $N_{\rm vox}$ is the total number of voxels (in Gigavoxels) and $V_{\rm Store}$ is the total data volume (in GB).  $M_{\rm GPU}$ is the mean memory per GPU in GB, which must be less than 8 GB so as not to exceed the memory bound of the NVIDIA GTX1080 graphics cards.   $T_{\rm Load}$ (in seconds) is the time measured for all of the spectral cubes to be loaded, rounded up to the nearest second.	 Statistical quantities calculated are the mean, sample standard deviation (SD), and median.}
\label{tbl:config}
\begin{tabular}{cccccc|ccc|ccc|cc}
	&		&		&		&		&		&	$\mathbf{N_{\rm vox}}$			&	$\mathbf{V_{\rm Store}}$			&	$\mathbf{M_{\rm GPU}}$			&	$\mathbf{N_{\rm vox}}$	&	$\mathbf{V_{\rm Store}}$	&	$\mathbf{M_{\rm GPU}}$	&	$\mathbf{T_{\rm Load}}$			&	$\mathbf{T_{\rm Load}}$	\\
Set	&	Config & Survey 	&	$\mathbf{N_{\rm W}}$ &	$\mathbf{N_{\rm T}}$	&	$\mathbf{N_{\rm L}}$	&	Mean	$\pm$	SD	&	Mean	$\pm$	SD	&	Mean	$\pm$	SD	&	Median	&	Median	&	Median	&	Mean	$\pm$	SD	&	Median	\\
\hline
A	&	2 $\times$ 2	&	W	&	20	&	-	&	-	&	0.61	$\pm$	0.06	&	2.43	$\pm$	0.23	&	0.49	$\pm$	0.05	&	0.61	&	2.45	&	0.49	&	19	$\pm$	2.9	&	21	\\	
	&		&	T	&	-	&	20	&	-	&	1.67	$\pm$	0.05	&	6.85	$\pm$	0.23	&	1.37	$\pm$	0.04	&	1.69	&	6.97	&	1.39	&	71	$\pm$	17.7	&	62	\\	
	&		&	L	&	-	&	-	&	20	&	0.41	$\pm$	0.13	&	1.63	$\pm$	0.52	&	0.32	$\pm$	0.11	&	0.38	&	1.51	&	0.30	&	21	$\pm$	8.5	&	18	\\	
	&		&	C	&	7	&	7	&	6	&	1.11	$\pm$	0.10	&	4.51	$\pm$	0.41	&	0.90	$\pm$	0.08	&	1.10	&	4.45	&	0.89	&	50	$\pm$	14.3	&	53	\\	\hline
B	&	4 $\times$ 2	&	W	&	40	&	-	&	-	&	1.23	$\pm$	0.06	&	4.91	$\pm$	0.24	&	0.98	$\pm$	0.05	&	1.24	&	4.98	&	1.00	&	38	$\pm$	2.5	&	38	\\	
	&		&	T	&	-	&	40	&	-	&	3.55	$\pm$	0.31	&	14.53	$\pm$	1.20	&	2.91	$\pm$	0.24	&	3.70	&	15.15	&	3.03	&	121	$\pm$	10.8	&	126	\\	
	&		&	L	&	-	&	-	&	40	&	0.84	$\pm$	0.02	&	3.37	$\pm$	0.08	&	0.67	$\pm$	0.02	&	0.83	&	3.33	&	0.67	&	40	$\pm$	3.5	&	40	\\	
	&		&	C	&	13	&	14	&	13	&	2.18	$\pm$	0.23	&	8.82	$\pm$	0.93	&	1.77	$\pm$	0.19	&	2.06	&	8.34	&	1.67	&	90	$\pm$	7.9	&	87	\\	\hline
C	&	4 $\times$ 4	&	W	&	80	&	-	&	-	&	2.41	$\pm$	0.04	&	9.65	$\pm$	0.17	&	1.93	$\pm$	0.04	&	2.40	&	9.61	&	1.92	&	73	$\pm$	3.6	&	74	\\	
	&		&	T	&	-	&	80	&	-	&	7.80	$\pm$	0.13	&	31.90	$\pm$	0.56	&	6.38	$\pm$	0.12	&	7.75	&	31.70	&	6.34	&	271	$\pm$	16.0	&	271	\\	
	&		&	L	&	-	&	-	&	80	&	1.58	$\pm$	0.00	&	6.33	$\pm$	0.00	&	1.27	$\pm$	0.00	&	1.58	&	6.33	&	1.27	&	57	$\pm$	3.5	&	57	\\	
	&		&	C	&	26	&	27	&	27	&	3.80	$\pm$	0.17	&	15.44	$\pm$	0.66	&	3.09	$\pm$	0.13	&	3.80	&	15.41	&	3.08	&	148	$\pm$	9.2	&	146	\\	\hline
D	&	6 $\times$ 4	&	W	&	120	&	-	&	-	&	3.63	$\pm$	0.03	&	14.51	$\pm$	0.11	&	2.90	$\pm$	0.02	&	3.64	&	14.55	&	2.91	&	105	$\pm$	2.1	&	104	\\	
	&		&	L	&	-	&	-	&	120	&	2.47	$\pm$	0.31	&	9.89	$\pm$	1.24	&	1.98	$\pm$	0.25	&	2.56	&	10.25	&	2.05	&	73	$\pm$	19.3	&	67	\\	
	&		&	C	&	40	&	40	&	40	&	5.61	$\pm$	0.22	&	22.85	$\pm$	0.94	&	4.57	$\pm$	0.19	&	5.68	&	23.21	&	4.64	&	194	$\pm$	11.8	&	197	\\	\hline
E	&	6 $\times$ 6	&	W	&	180	&	-	&	-	&	5.42	$\pm$	0.03	&	21.67	$\pm$	0.13	&	4.34	$\pm$	0.03	&	5.43	&	21.71	&	4.34	&	156	$\pm$	7.0	&	159	\\	
	&		&	L	&	-	&	-	&	180	&	3.20	$\pm$	0.24	&	12.79	$\pm$	0.97	&	2.56	$\pm$	0.20	&	3.27	&	13.08	&	2.62	&	95	$\pm$	2.7	&	96	\\	
	&		&	C	&	60	&	60	&	60	&	8.35	$\pm$	0.25	&	33.94	$\pm$	1.05	&	6.79	$\pm$	0.21	&	8.46	&	34.34	&	6.87	&	285	$\pm$	11.2	&	281	\\	\hline
\end{tabular}
\end{table*}

We categorise the survey data products in terms of: (1) the number of sources ($N_{\rm s}$) in each survey catalogue; (2) the typical dimensionality of the data cubes (measured as spatial or spectral pixels); (3) the number of voxels (in Megavoxels or Mvox);  and (4) the storage size (in Megabytes or MB) for an individual cube.  For all three datasets, the spectral cubes were stored (and loaded into {\tt encube}) using the Flexible Image Transport System (FITS) format \citep{Wells81,Hanisch01,Pence10}. See Table \ref{tbl::surveys} for further details, where we present the minimum, maximum and median values for the dimensions, voxel counts and storage sizes for the WHISP, THINGS and LVHIS catalogues.  

To simplify both the benchmarking investigation and VDAR evaluation, we make several minor modifications to the datasets in their published forms:
\begin{itemize}
    \item WHISP: Initial inspection of a sub-set of WHISP galaxies revealed that many of the spectral cubes have high levels of flux (relative to the peak source flux) at either end of the spectral band.  Rapid identification of such systematic effects is an example of the type of SIMD quality control activity that comparative visualisation can address (see Section \ref{sct:quality}).  For all of the WHISP cubes, we created new FITS files where we set the data values in the first eight and last eight spectral channels to zero.  This does not change the load times for the mock surveys but does improve the default visualisation via texture-based volume rendering.
    \item THINGS: We did not use the spectral cube for NGC 3031 (M81) in our benchmarking. As NGC 3031 is a nearby grand design spiral in Ursa Major, the spectral cube is much larger than other galaxies in the sample with $2201 \times 2201$ spatial and $178$ spectral channel pixels.  The file size of 3.45 GB is approximately half of the available memory on a GTX1080 GPU.  Such a large source would not be typical of new extragalactic sources discovered with blind surveys.
    \item LVHIS: A spectral cube data for NGC 5128 (LVHIS 048) was not available from the survey web-site, and we note a replication of data between sources LVHIS 014 and LVHIS 016, which are both identified as the dwarf irregular galaxy AM 0319-662. Removing LVHIS 016 and LVHIS 048 from the samples leaves us with $N_{\rm s} = 80$.
\end{itemize}

\section{Benchmarking comparative workflows}
\label{sct:benchmarks}
In this section, we report on benchmarking activities undertaken with the implementation of {\tt encube} on the Swinburne Discovery Wall.

\subsection{Benchmarks}
Previous system benchmarks reported in \citet{Vohl16} were performed with the Monash CAVE2$^{\rm TM}$. For deployment on the Swinburne Discovery Wall, we report: (1) the total (i.e. parallel) load time, $T_{\rm Load}$, for a configuration displaying $N_{\rm cube}$ spectral cubes; and (2) the steady-state minimum frame rate, $F_{\rm rate}$, in frames/second.    We consider both the frame rate per column, looking for variations in performance, along with the overall mean, standard deviation, and median of $F_{\rm rate}$.  

Frame rate quantities are calculated from the {\tt S2PLOT} displays on columns 2 to 5 (see Figure \ref{fig:sdw}). Column 1 is used for additional management and coordination tasks, and in order to access the user interface in the web browser, the {\tt S2PLOT} display is not resized over both 4K-UHD monitors.  The higher $F_{\rm rate}$ values reported for column 1 show the overall reduced graphics workload when data is visualised on one 4K-UHD monitor instead of two.

We obtained a total of 54 independent benchmarks for five different configurations (Sets A--E), displaying $N_{\rm cube}$ = 20, 40, 80, 120 or 180 spectral cubes in total using the per-column configurations summarised in Table \ref{tbl:config}.   The main limiting factors on $N_{\rm cube}$ are the available GPU memory (8 GB/GPU for each of the five NVIDIA GTX1080 GPUs of the Swinburne Discovery Wall) and the number of columns of monitors.  A simple upgrade path to improve performance is to replace these five older-generation GPUs with higher-memory alternatives.

The benchmark configurations were generated comprising either spectral cubes from a single survey (denoted as [W]HISP, [T]HINGS or [L]VHIS) or from the combination of the three input surveys (denoted as [C]ombination). For scenarios where $N_{\rm cube}$ exceeds the survey size, $N_{\rm s}$ (see Table \ref{tbl::surveys}), random sampling with replacement is used to generate an appropriately-sized data set. For the combination survey, random sampling with replacement is used to generate a mock survey that is roughly equally split between the three input catalogues.   

\begin{figure*}[ht]
    \centering
    \includegraphics[width=17cm]{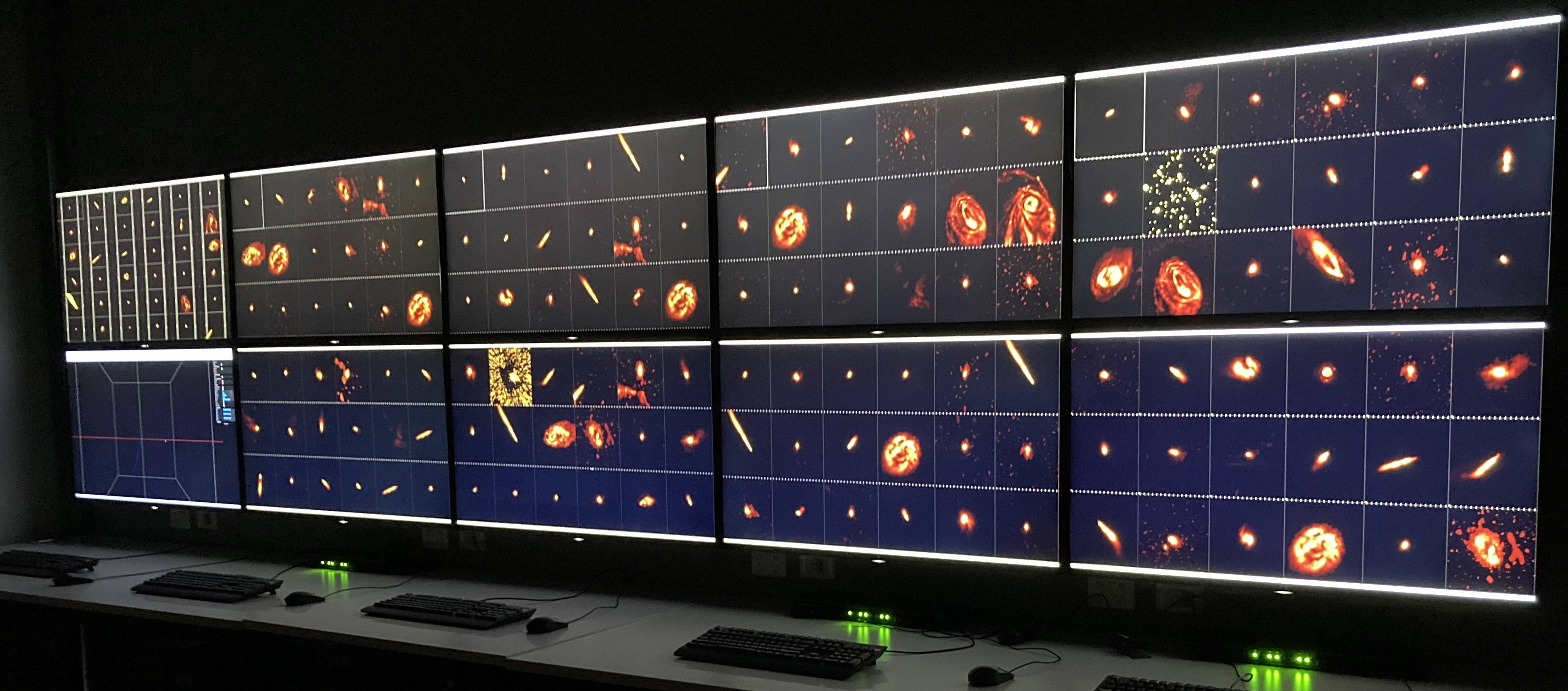}
    
    \includegraphics[width=17cm]{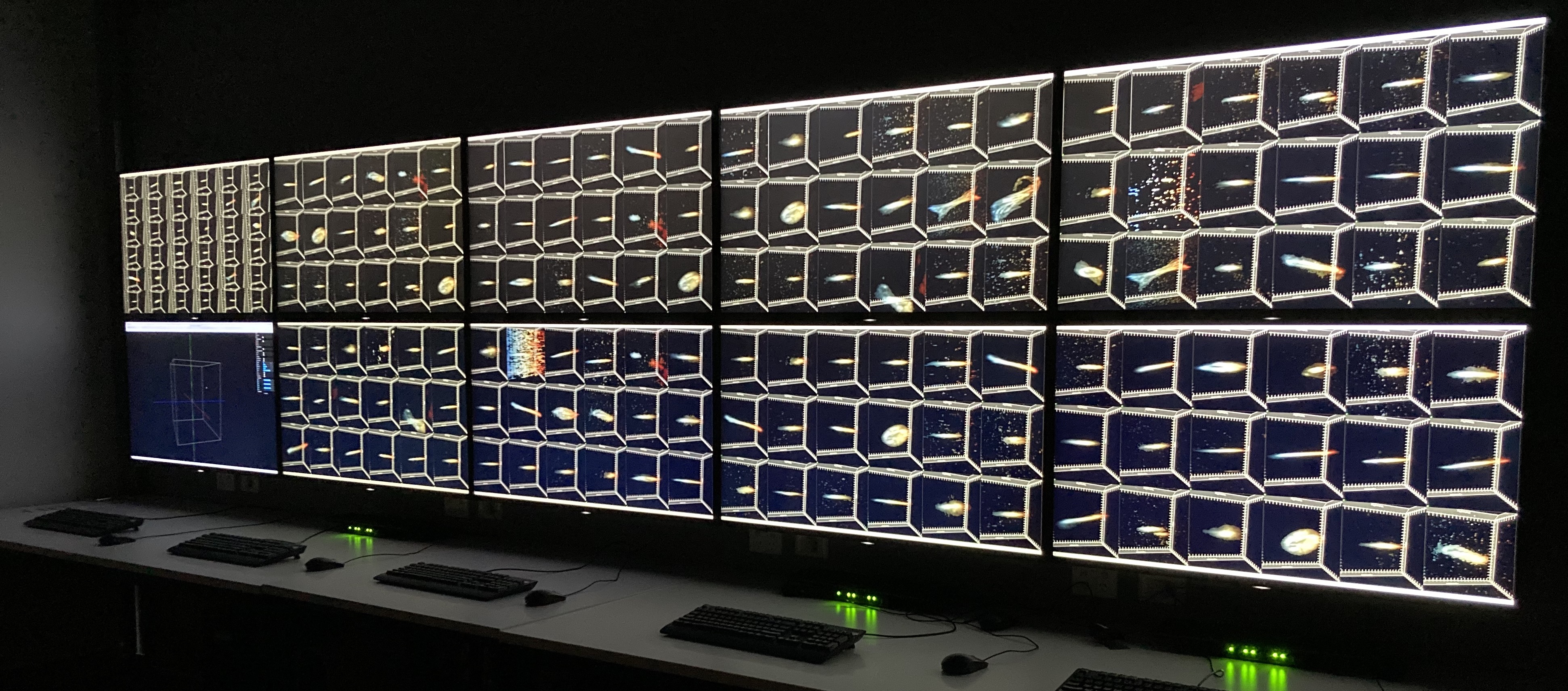}
    \caption{Simultaneous visualisation of 180 spectral cubes from the LVHIS H{\sc i} survey. Sources are randomly sampled with replacement, resulting in repetition of objects across the display.  This configuration loads in less than 100 seconds.  (Top) A zoomed-in view in showing the spatial distribution of H{\sc i} using a heat-style colour map where low signal is black and  high signal is white. (Bottom) All cubes are rotated to show the kinematic structure along the spectral axis. A blue-red two-ended colour map is used to aid with identifying H{\sc i} that is either blue-shifted or red-shifted with regards to the observer. }
    \label{fig:LVHIS-180}
\end{figure*}

Figure \ref{fig:LVHIS-180} demonstrates the use of the two different colour-mapping methods for a mock LVHIS survey with 180 spectral cubes.  The top panel uses a heat-style colour map, while the bottom map colours based on the relative velocity with respect to the middle spectral channel, which is assumed to be the kinematic centre.

\begin{table}[ht]
    \caption{With spectral cube data stored in the FITS format, there is a slight variation in the ratio between the total data volume, $V_{\rm Store}$ measured in GB, and the number of voxels, $N_{\rm vox}$ measured in Gigavoxels across all 54 survey configurations.  This is due, in part, to the varying lengths of the FITS headers.   }
    \label{tbl:storevox}
    \centering
    \begin{tabular}{c|c}
{\bf Quantity} & 
    $\mathbf{V_{\rm Store}/N_{\rm vox}}$ \\
    \hline
    Mininum & 3.97 \\
    First quartile & 4.00 \\
    Median & 4.02 \\
    Mean & 4.03 \\
    Third quartile & 4.06 \\
    Maximum & 4.12 \\
    \hline
    \end{tabular}
\end{table}

\begin{figure*}[ht]
    \centering
    \includegraphics[width=8cm]{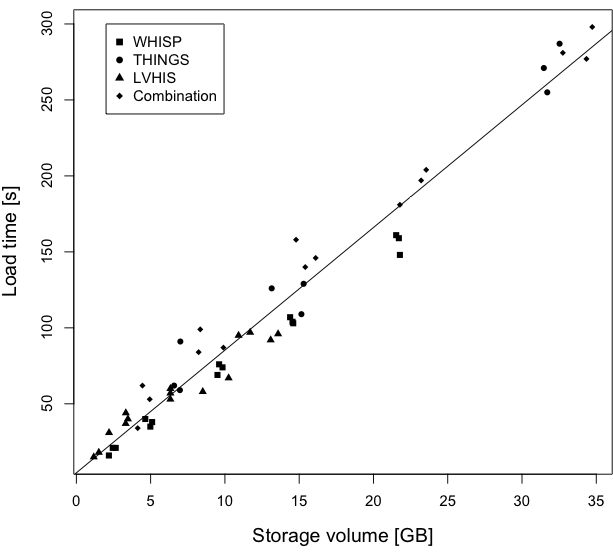} \hspace{1cm}
    \includegraphics[width=8cm]{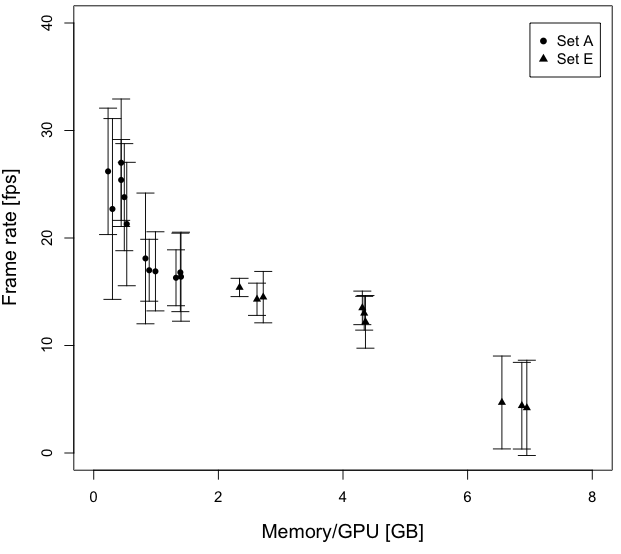}
    \caption{(Left panel) Based on the 54 independent benchmarks (see the summary in Table \ref{tbl:config}), the total time taken to load all spectral cubes for a given input configuration grows linearly with the storage volume.  Load times are rounded up to the nearest second. Symbols are used to denote the four different input surveys; WHISP (square), THINGS (circle), LVHIS (triangle) or Combination (diamond). (Right panel) From a subset of 21 benchmarks, the minimum recorded frame rate decreases as the mean memory per GPU of the Discovery Wall increases.  Plotted values are the mean $\pm$ standard deviation of the minimum observed frame-rate across columns 2-5 of the Discovery Wall (see Table \ref{tbl:framerate}).  Frame rate benchmarks were only obtained for Set A (circle) and Set E (triangle), with $N_{\rm cube}$ = 20 or 180 respectively.  A reasonable frame-rate for interactivity is above 10 frames/sec, which was achieved except in the Combination configuration containing higher-data volume THINGS spectral cubes.  }
    \label{fig:benchgraph1}
\end{figure*}

To mitigate the impact of memory caching on measurements of $T_{\rm Load}$, we generated three independent combinations of spectral cubes for each of the W, T, L and C configurations. A single benchmark value of $T_{\rm Load}$ was obtained for each of the three alternatives, along with the measurements of $F_{\rm rate}$.   For the 80-cube instance, we note that all LVHIS cubes are used, but they are randomly assigned between the five columns of the Discovery Wall for each benchmark instance.  

We did not generate configurations with $N_{\rm T} > 80$ as these data volumes exceed the memory capacity of the GPUs.  The THINGS galaxies are the highest-resolution spectral cubes considered in this study, and are not as representative of the typical resolved or partially-resolved new detections that will arise from ASKAP or MeerKAT H{\sc i} surveys.

Due to the presence of differing numbers of key-value pairs in the FITS headers, there is slight variation (see Table \ref{tbl:storevox}) in the ratio between \textbf{$V_{\rm Store}$} (the total data volume in GB) and $N_{\rm vox}$ (the total number of voxels in Gigavoxels) for the 54 independent survey configurations.   The result of a least-squares fit to the these two quantities was:
\begin{equation}
    V_{\rm Store} = 4.07 N_{\rm vox}- 0.084 \; \mbox{GB},
\end{equation}
with the mean and sample standard deviation between measured and modelled values for $V_{\rm Store}$ calculated to be $-9.4 \times 10^{-6}$ GB and $0.13$ GB respectively.   For simplicity, we can approximate $V_{\rm Store} \sim 4 N_{\rm vox}$ as expected for a data format using four bytes per voxel.

\subsection{Procedure}
All of the spectral cubes are stored on the workstation associated with column 1 of the Swinburne Discovery Wall (the Master Node - see Figure \ref{fig:layout}), and the other workstations access this data through a network file sytem (NFS) mount (see \ref{sub:keycomp}).  Consequently, we expect that the limiting factors on $T_{\rm Load}$ are: (1) the network bandwidth between each Process and Render workstation and the Master; (2) the read time from the NFS-mounted drive; and (3) the processing overheads due to pre-computation of statistical parameters, as noted at the end of \ref{sub:keycomp}.

The following procedure was used to conduct each of the benchmark trials:
\begin{enumerate}
    \item The set of spectral cubes is randomly selected either without replacement (when $N_{\rm cube} \leq N_{\rm s}$) or with replacement, and a database file is generated in the comma-separated variable (CSV) format required by {\tt encube}.
    \item Symbolic links are generated to each of the $N_{\rm cube}$ spectral cubes, to minimise the duplication of data on the Master workstation.
    \item Modifications to the {\tt encube} configuration file (keyword-value pairs using JavaScript Object Notation\footnote{JSON: https://www.json.org/json-en.html}) are made, specifically the number of rows and columns of {\tt S2PLOT} panels per column of the Discovery Wall, the total number of panels per workstation, and the names of the workstations.
    \item {\tt Encube} is launched from the Master workstation using the JSON configuration file, with calls to start the software on the Process and Render nodes.  Socket connections are established between the Master and the Process and Render nodes, and a port is opened for connection to the user interface (UI).
    \item The {\tt encube} UI is activated as a web-page in the Firefox browser on the Master machine.  The UI displays the database of spectral cube files.  The required files are selected and timing for $T_{\rm Load}$ commences on mouse-clicking the Load button.
    \item Timing ends when all spectral cubes are displayed.  As timing is performed by hand, all times are rounded up to the nearest whole second to account for the timekeeper's reaction time.
    \item For the subset of configurations where frame rates are also recorded on a per-column basis, an autospin signal is triggered from the UI which causes all of the spectral cubes to rotate around the vertical axis.  At each of the five keyboards attached to the columns (see Figure \ref{fig:sdw}), the {\tt d} key is pressed, activating the {\tt S2PLOT} graphics debug mode, which reports the instantaneous frame rate (measured over a moving window of 5 seconds duration).  After each spectral cube has completed several complete rotations, the lowest measured frame rate is recorded.  This presents the worst-case scenario, as the frame rate is a strong function of both the viewing angle of a spectral cube and the fraction of the screen that is mapped to data voxels.
    \item Once benchmark quantities have been recorded, a signal to stop the {\tt encube} instances is initiated from the UI, and all of the processes are stopped from the Master workstation.  It takes approximately 60 seconds for all nodes to release their socket connections ready for the next full iteration of the procedure.
\end{enumerate}

\begin{table*}[ht]
    \caption{Indicative frame rates for each of the five columns of the Swinburne Discovery Wall using a subset of the survey configurations.   Quantities and units not defined elsewhere (see the caption to Table \ref{tbl:config}) are the version number of each mock survey, Ver, and the lowest measured column-based frame rates, $F_i$, in frames/s, recorded after several complete rotations of each spectral cube. Subscripts 1--5 on the frame rate indicate the column of the Discovery Wall, numbered from left to right as seen in Figure \ref{fig:sdw}.}
    \label{tbl:framerate}
    \centering
    \begin{tabular}{cccrrrrrrrrrc}
$\mathbf{N_{\rm cube}}$	&	{\bf Survey}	&	{\bf Ver} &	$\mathbf{N_{\rm vox}}$	&	$\mathbf{V_{\rm Store}}$	&	$\mathbf{M_{\rm GPU}}$	&	$\mathbf{F_1}$	&	$\mathbf{F_2}$	&	$\mathbf{F_3}$	&	$\mathbf{F_4}$	&	$\mathbf{F_5}$	&	{\bf Mean}	$\mathbf{\pm}$	{\bf SD}	&	{\bf Median}	\\
\hline
20	&	W	&	1	&	0.61	&	2.45	&	0.49	&	30.0	&	29.8	&	22.6	&	17.8	&	24.8	&	23.8	$\pm$	5.0	&	23.7	\\
20	&	W	&	2	&	0.55	&	2.19	&	0.44	&	30.0	&	29.6	&	27.4	&	22.6	&	21.8	&	25.4	$\pm$	3.8	&	25.0	\\
20	&	W	&	3	&	0.66	&	2.65	&	0.53	&	30.0	&	29.9	&	18.0	&	18.2	&	19.2	&	21.3	$\pm$	5.7	&	18.7	\\
20	&	T	&	1	&	1.71	&	7.00	&	1.40	&	30.0	&	22.5	&	13.2	&	15.0	&	15.0	&	16.4	$\pm$	4.1	&	15.0	\\
20	&	T	&	2	&	1.69	&	6.97	&	1.39	&	30.0	&	22.3	&	15.0	&	15.0	&	15.0	&	16.8	$\pm$	3.7	&	15.0	\\
20	&	T	&	3	&	1.61	&	6.58	&	1.32	&	30.0	&	20.2	&	15.0	&	15.0	&	15.0	&	16.3	$\pm$	2.6	&	15.0	\\
20	&	L	&	1	&	0.38	&	1.51	&	0.30	&	30.0	&	30.0	&	15.0	&	30.0	&	15.9	&	22.7	$\pm$	8.4	&	23.0	\\
20	&	L	&	2	&	0.29	&	1.17	&	0.23	&	30.0	&	30.0	&	27.2	&	17.6	&	30.0	&	26.2	$\pm$	5.9	&	28.6	\\
20	&	L	&	3	&	0.55	&	2.20	&	0.44	&	30.0	&	29.9	&	30.0	&	30.0	&	18.1	&	27.0	$\pm$	5.9	&	30.0	\\
20	&	C	&	1	&	1.22	&	4.94	&	0.99	&	30.0	&	22.4	&	15.0	&	15.1	&	15.0	&	16.9	$\pm$	3.7	&	15.1	\\
20	&	C	&	2	&	1.02	&	4.13	&	0.83	&	30.0	&	27.2	&	15.0	&	15.1	&	15.0	&	18.1	$\pm$	6.1	&	15.1	\\
20	&	C	&	3	&	1.10	&	4.45	&	0.89	&	30.0	&	21.1	&	16.7	&	15.0	&	15.0	&	17.0	$\pm$	2.9	&	15.9	\\
180	&	W	&	1	&	5.43	&	21.71	&	4.34	&	18.5	&	15.3	&	11.7	&	12.4	&	12.7	&	13.0	$\pm$	1.6	&	12.6	\\
180	&	W	&	2	&	5.44	&	21.78	&	4.36	&	18.1	&	15.7	&	11.7	&	11.4	&	10.0	&	12.2	$\pm$	2.5	&	11.6	\\
180	&	W	&	3	&	5.38	&	21.53	&	4.31	&	18.0	&	15.5	&	13.9	&	12.1	&	12.4	&	13.5	$\pm$	1.6 &	13.2	\\
180	&	L	&	1	&	2.93	&	11.70	&	2.34	&	19.6	&	16.7	&	15.0	&	15.0	&	15.0	&	15.4	$\pm$	0.9	&	15.0	\\
180	&	L	&	2	&	3.27	&	13.08	&	2.62	&	23.1	&	15.4	&	12.1	&	15.0	&	14.8	&	14.3	$\pm$	1.5	&	14.9	\\
180	&	L	&	3	&	3.40	&	13.58	&	2.72	&	15.0	&	16.9	&	11.2	&	15.0	&	15.0	&	14.5	$\pm$	2.4	&	15.0	\\
180	&	C	&	1	&	8.53	&	34.73	&	6.95	&	7.6	&	1.8	&	1.6	&	2.5	&	10.8	&	4.2	$\pm$	4.4	&	2.2	\\
180	&	C	&	2	&	8.46	&	34.34	&	6.87	&	3.9	&	3.3	&	1.9	&	2.1	&	10.4	&	4.4	$\pm$	4.0	&	2.7	\\
180	&	C	&	3	&	8.07	&	32.75	&	6.55	&	10.4	&	2.0	&	4.7	&	1.3	&	10.8	&	4.7	$\pm$	4.3	&	3.4	\\
\hline
    \end{tabular}
\end{table*}

The outcomes of the benchmarks are reported as follows:
\begin{itemize}
    \item A statistical summary (mean, sample standard deviation, and median) of $T_{\rm Load}$ for the three independent instances of each survey configuration is presented in the final two columns of Table \ref{tbl:config}.
    \item The survey load time is plotted as a function of the storage volume in the left-hand panel of Figure \ref{fig:benchgraph1}.  All 54 independent benchmarks for $T_{\rm Load}$ are presented, with symbols for WHISP (squares), THINGS (circles), LVHIS (triangles) and the Combination survey (diamonds).
    \item Individual values, and statistical characterisation of $F_{\rm rate}$ is presented in Table \ref{tbl:framerate}. A subset of 21 configurations was considered here: Set A, with $N_{\rm cube}$ = 20 and Set E, with $N_{\rm cube}$ = 180.
    \item The minimum frame rates for each of columns 2-5 for Set A (circles) and Set E (triangles) is plotted in the right-hand panel of Figure \ref{fig:benchgraph1} as a function of the mean memory per GPU on the Discovery Wall.
\end{itemize}

A linear relationship exists between $T_{\rm Load}$ (s) and $V_{\rm Store}$ (GB), with a least squares fit result:
\begin{equation}
    T_{\rm Load} = 8.07 V_{\rm Store} + 4.58 \; \mbox{seconds}.
    \label{eqn:loadstore}
\end{equation}
The mean and sample standard deviation between measured and modelled values for $T_{\rm Load}$ were calculated to be $5.6 \times 10^{-4}$ seconds and $13.9$ seconds respectively.    The Pearson correlation coefficient between $T_{\rm Load}$ and $V_{\rm Store}$ was $r = 0.98$.  For completeness, we find:
\begin{equation}
    T_{\rm Load} = 32.83 N_{\rm vox} + 4.063 \; \mbox{seconds}
    \label{eqn:loadvox}
\end{equation}
with $N_{\rm vox}$ in Gigavoxels.

We discuss the implications of our benchmarking activities in Sections \ref{sct:load} to \ref{sct:through}.  In the next section, we provide details of our VDAR evaluation.

\section{Visual data analysis and reasoning study}
\label{sct:vdar}
\citet{Lam12} \citep[and see also][]{Isenberg13} proposed a taxonomy for understanding and evaluating visualisation methods. We select the VDAR approach to examine typical survey-scale discovery-based research processes, relevant for current and future extragalactic H{\sc i} surveys.

VDAR includes methodologies for evaluating the effectiveness or efficacy by which a visualisation tool helps to generate domain-specific actionable knowledge or understanding.  VDAR methods, which often are based on case studies, investigate ``the tool used in its intended environment with realistic tasks undertaken by domain experts'' \citep{Lam12}, with an emphasis on the process rather than measurements of outcomes. 

Our user group for the VDAR study comprises only the authors of this work.  This cohort includes domain experts (i.e. H{\sc i} astronomers with relevant experience in the observation, analysis and visualisation of spectral cubes), as required with the VDAR methodology. We assert that these experiences are representative of the broader H{\sc i} research community.  

Alternative evaluation methodologies for visualisations and visualisation systems \citep{Lam12,Isenberg13} that we did not pursue include Evaluating Collaborative Data Analysis (CDA), which focuses on the process of collaboration and how it is supported by a visualisation solution, and User Performance (UP), which uses controlled experiments to measure, for example, the time taken for different users to complete tasks. As a point of comparison, \citet{Meade14} used the UP methodology to measure task performance when novice and expert participants completed an object identification activity using either a standard desktop monitor or a TDW.

To provide relevant scenarios for the VDAR study, we 
consider three important SIMD processes that may be required when analysing extragalactic H{\sc i} survey data:  (1) {\em quality control} of individual candidate spectral cubes; (2) {\em candidate rejection}, whereby false-positive detections from automated source finders are rejected; and (3) {\em morphological classification}, identifying and sorting sources into categories based on observed structural or kinematic properties.  These three processes currently require some level of visual inspection [which may include the use of either projected moment maps or 3D visualisation methods, depending on the workflow preferences of the researcher(s) involved] in order to produce reliable, science-ready catalogues from large-scale, next-generation surveys. 

It is important to note that our VDAR study does not intend to demonstrate new knowledge about any of the three input H{\sc i} surveys -- WHISP, THINGS, and LVHIS -- as all have been well-studied in many other contexts.  They stand in as proxies for future H{\sc i} survey data products that are, potentially, being viewed for the very first time by members of the research team.  As such, there may be unexpected, or unexplained, features that are present in the data products, necessitating appropriate follow-up actions once they have been identified.

Alternatively, the comparative visualisation stage may reveal that all is well with automated calibration or processing steps (e.g. model-fitting) at an early stage of science operations, thus serving its purpose.  For a related example where the use of an alternative display technology evolves throughout the lifetime of an astronomical research project, see Section \ref{sct:evolution}.

\begin{figure}[ht]
    \centering
\includegraphics[width=\textwidth]{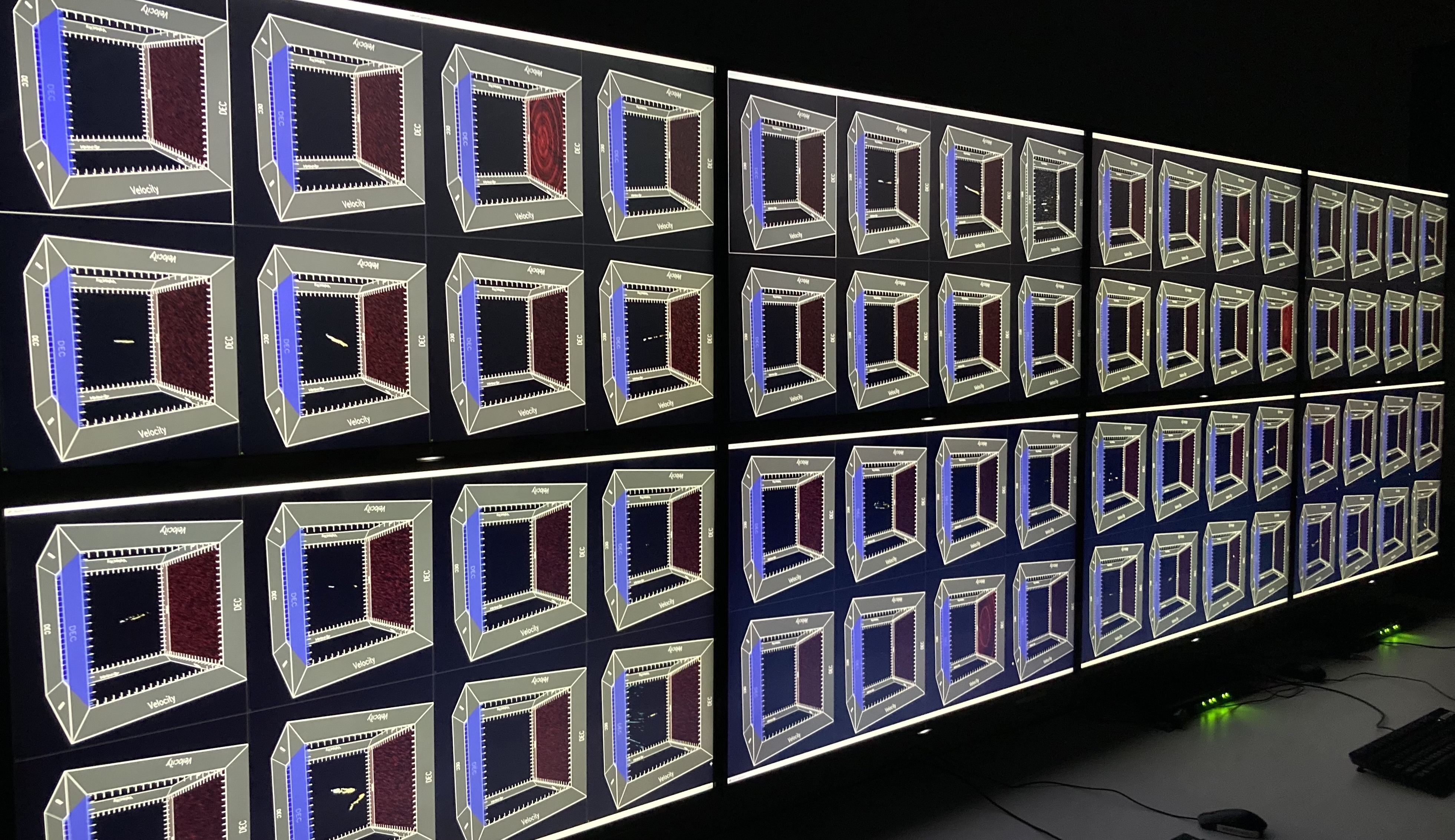}
\includegraphics[width=\textwidth]{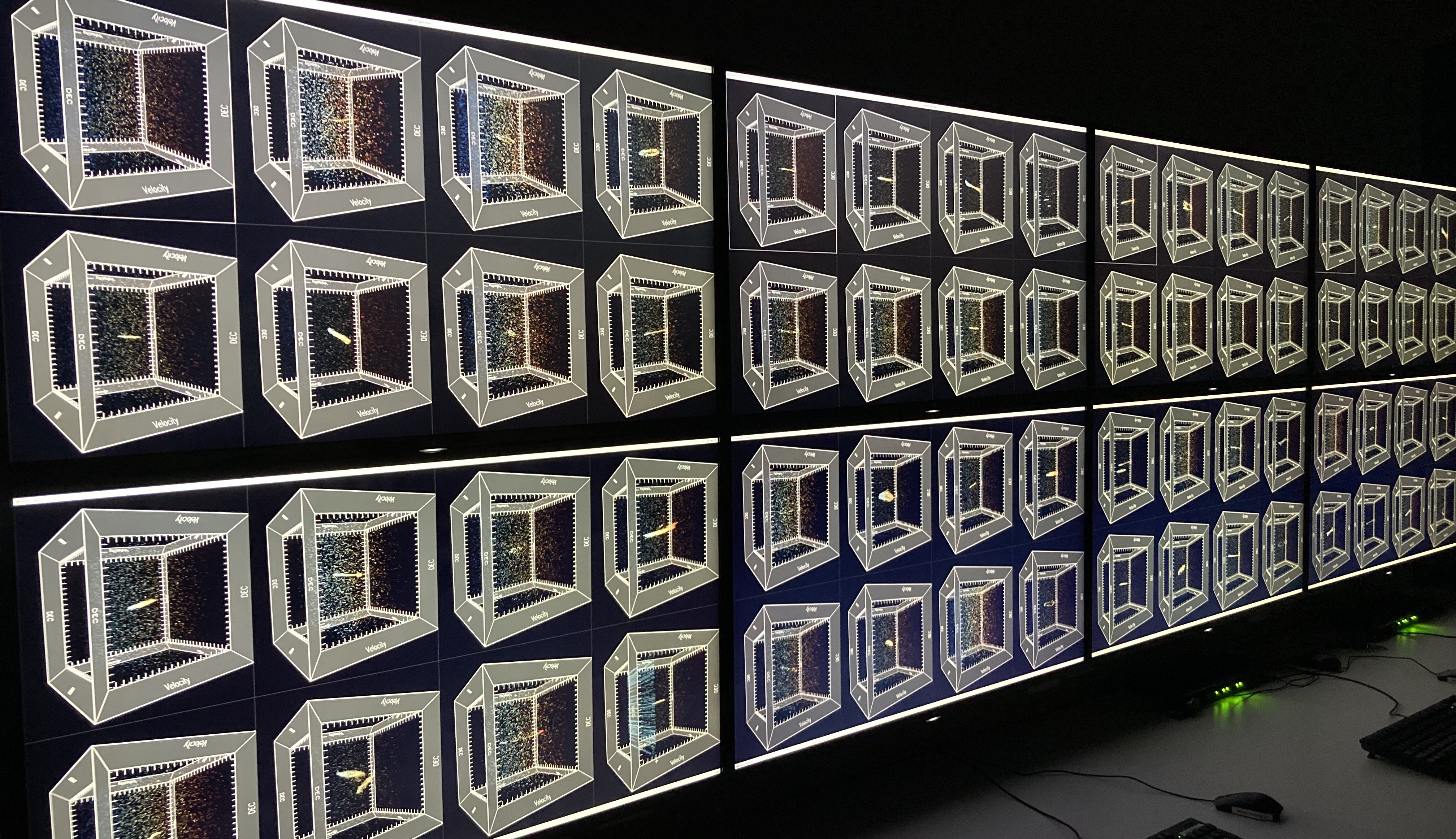}
    \caption{A quality control activity using {\tt encube} and the Swinburne Discovery Wall to visualise 80 WHISP spectral cubes.  (Top) Visualisation of the mock survey using the data as obtained from the WHISP survey website. We observe that the volume rendering has not worked as expected. In 77 cubes, there is visible excess flux at both ends of the spectral axis. This is seen as the strong blue and red features in each cube, making it difficult to see the WHISP galaxies in most cases. (Bottom) By choosing to reset data values to zero in the first eight and last eight channels of each WHISP spectral cube, the kinematic H{\sc i} structures are now visible.  }
    \label{fig:qc}
\end{figure}

\subsection{Quality control}
\label{sct:quality}
When an H{\sc i} source finding pipeline is applied to a large-scale survey cube, the output is a set of individual source cubelets.  Prior to their use in further analysis, there is value in performing by-eye quality control, to ensure that there are no significant issues with the data quality.   This step would be expected to include looking for: (1)  bad channels; (2) calibration errors such as poor continuum subtraction; (3) objects that have not been correctly extracted, such as extended sources that exceed the boundaries of the extracted cubelet; and (4) radio frequency interference.

The VDAR study we performed to understand the quality control process relates to our observation when first visualising a sub-set of WHISP galaxies with {\tt encube}.  As noted in Section \ref{sct:surveydata}, spectral channels at both ends of the band-pass contain excess flux.  We illustrate this issue in the top panel of Figure \ref{fig:qc}, using an 80-cube configuration.   The excess flux is visible in 77 of the cubes displayed. This is seen as the strong blue and red features in each cube, making it difficult to see the WHISP galaxies themselves.

With {\tt encube}, it is immediately clear that a quality control issue is present and is impacting a sizeable portion of the survey.    From Table \ref{tbl:config}, it takes less than 90 seconds to load the 80 WHISP cubes, and then less than 60 seconds to identify the 3 cases that do not appear to be affected.   Performing this task in a serial fashion would require individual loading and inspection of spectral cubes: it would take much longer than 150 seconds to determine the extent of the quality control issue in order to take an appropriate action.

Our solution was to replace data values in the first eight and last eight channels of each WHISP spectral cube. This has the desired effect, revealing the kinematic structures of the sources (see the lower panel of Figure \ref{fig:qc}).

There will be an additional quantity of time required to resolve any quality control issue.  In this case, we needed to write and execute a C-language program using the {\tt CFITSIO}\footnote{https://heasarc.gsfc.nasa.gov/fitsio/} \citep{Pence99} library to create modified FITS-format data cubes for the WHISP galaxies.  For a future H{\sc i} survey, it may require modification or re-tuning of an automated calibration pipeline. However, this time is independent of whether the quality control visualisation is approached in a serial or parallel fashion. Indeed,  comparative visualisation provides a more rapid demonstration that the intervention had the desired effect.

Our approach to comparative quality control with {\tt encube} is consistent with the model of sensemaking presented by \citet{Pirolli05}.  Here, our use of the Discovery Wall has two dimensions: (1) a {\em foraging loop}, organising data, searching for relations, and gathering evidence; and (2) a {\em sensemaking loop}, where alternative hypotheses are posed and examined, leading to a presentation of the outcomes.

In the foraging loop, we determine that a quality control issue exists, as the initial volume renderings are not consistent with the expected profiles of H{\sc i}-detected sources.  This issue impacts a significant number of spectral cubes in the sample (77 out of 80). Through physical navigation (i.e. moving to different locations near the Discovery Wall), the viewer can change their attention from a single object to an ensemble in order to gather evidence regarding the possible cause of the failed visualisations.  

In the sensemaking phase, we decide that a first course of action is to remove the impact of the excess flux in all spectral cubes, and visualise the outcomes.  Further investigation could include selecting the subset of those spectral cubes most strongly impacted, in order to determine the cause(s) of the excess flux.

\subsection{Candidate rejection}
\label{sct:candidate}
An unwanted outcome of automated source finders is the generation of false-positive detections.  This is particularly true in their early phase of operation of new survey programs, when source finders may not have been tuned optimally to the specific characteristics of the data.  But false-positives may persist throughout the lifetime of a survey. 

One way to improve the accuracy of source-finders is to raise the acceptance threshold, so that fewer candidates make it through the processing pipeline for further inspection and analysis.   This approach reduces the discovery space, with many interesting objects remaining undetected.   By lowering the acceptance criteria, more false candidates will need to be reviewed and ultimately rejected. This can be a particularly labour intensive phase. 

Visual inspection is the simplest way to distinguish between true sources and false detections, but may require an appropriate level of expertise.  Here, again, quality control processes will be crucial, as individual cubelets may suffer from anomalies from processing, calibration, or interference.

Our bespoke visualisation environment permits rapid inspection and comparison of many  sources at the same time, improving the way that decisions are made regarding the nature of candidates.   The VDAR study we performed to understand the candidate rejection process was to:
\begin{itemize}
    \item Load one of the 80-cube combination surveys (Set C), with $T_{\rm Load} \sim 150$ seconds.  The combination survey includes a high proportion of spatially resolved galaxies from the THINGS and LVHIS catalogues.
    \item Visually inspect every source, looking for the spatially resolved galaxies, and then identifying which of these did not immediately match the expected template of a grand design spiral galaxy.   
\end{itemize}

It took less than three minutes to visually inspect all 80 cubes. While some resolved, non-spiral galaxies were very easy to identify, others require additional time in order to reach a decision. Here, the use of the volume rendering technique allows for individual sources, or sets of sources, to be rotated such that either the spatial or kinematic structure can be used to reach a decision.

\begin{figure*}[ht]
\includegraphics[width=19cm]{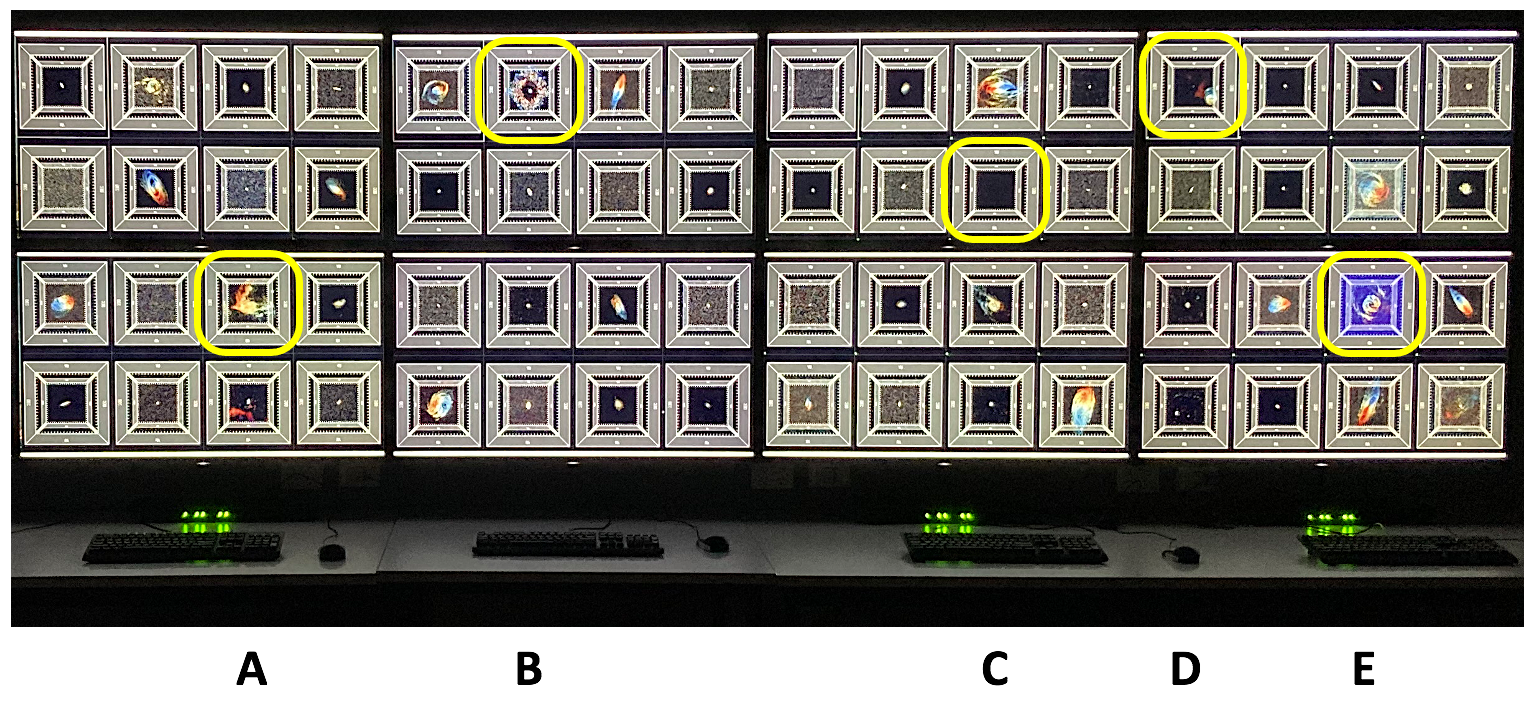}
\caption{A demonstration of {\tt encube} in use for a SIMD candidate rejection or morphological classification activity.  Shown here are columns 2-5 of the Swinburne Discovery Wall. Five sources of interest (labelled A-E under the column in which they are located, and described in Section \ref{sct:candidate}) have been highlighted for further investigation.  The overview provided by visualising many small-multiples allows for rapid identification of these five sources, which show spatial or spectral features that are quite different to the other 75 sources in the survey sample.}
\label{fig:morphoclass}
\end{figure*}

Figure \ref{fig:morphoclass} shows columns 2--5 of the Swinburne Discovery Wall, with labels under the image used to identify five sources of interest (A-E):
\begin{enumerate}
    \item Source A (THINGS, NGC 3077) is spatially resolved, but shows a disrupted H{\sc i} structure.  NGC 3077 is connected to a larger neighbouring spiral galaxy, M81, by an H{\sc i} bridge \citep{vdHulst79};
    \item  Source B (LVHIS, ESO 245-G007) shows a ``tube-like'' feature  (readily apparent when rotating the spectral cube) surrounding a central, somewhat spatially unresolved object;
    \item For source C (WHISP, UGC01178), there is no visible flux, which is likely due to a poor choice of the default visualisation parameters;
    \item Source D (LVHIS, AM 0319-662) comprises two H{\sc i} detections, with the more prominent source offset from the centre of the cube. The central LVHIS source is a dwarf irregular galaxy, a companion to NGC 1313 at the lower right of the cube \citep{Koribalski2018MNRAS.478.1611K}; and
    \item Source E (THINGS, NGC5236) is a spiral galaxy, but the overall blue feature extending across the source indicates some additional processing may be required. In particular, this can be explained as this source, Messier 83, is known to have an H{\sc I} diameter much larger than the VLA primary beam with which it was observed in the THINGS project.  The overview provided by many small-multiples rapidly highlight this source's distinctive feature, which was not present in any of the other 79 sources in this sample. 
\end{enumerate}
Identification of these five  ``anomalous'' cases occurs rapidly, when the viewer is able to both see a large sample (i.e. comparative visualisation, by stepping back from the Discovery Wall) and investigate an individual object in more detail (by moving closer to view, or interact with, an object of interest).

To close the loop on candidate rejection, a minor modification to {\tt encube} would allow each spectral cube to be tagged in real time as a true or false detection, which would then be fed back to the source finder to improve the true detection rate.

\subsection{Morphological classification}
Once a catalogue of robust detections has been gathered, the nature of the sources must be considered.   For previously known objects, a morphological classification has likely already occurred.  For new discoveries, an initial classification can be provided.  

For future H{\sc i} surveys conducted with wide-field interferometric imaging, the extended structure of many sources will be visible. This includes detecting the presence of low column density features such as bridges, tails, etc. Consequently, visual morphological classification of complete, unbiased, sub-populations of sources will be possible.  Indeed, with a statistically significant population of H{\sc i} galaxies, selected in an unbiased (i.e. blind survey) fashion, it becomes possible to develop new morphological categories  -- beyond the standard Hubble classification -- that may correlate with the local or global environment or integral properties, such as the H{\sc i} mass.

The morphological classification process shares many similarities with the candidate rejection phase, and we appeal to the same VDAR study as in Section \ref{sct:candidate}.  The two features of our bespoke visualisation environment that provide an alternative approach to morphological classification, at scale, are: (1) the use of volume rendering, which allows each spectral cube to be rotated around any axis, providing immediate access to both spatial and kinematic information; and (2) the comparative nature of the display configuration, which makes it easy to go back-and-forth between specific objects in order to reach a decision regarding the classification. This might mean a change in the outcome of an initial or even pre-existing classification, or the recognition that a new sub-class of objects had been identified.

\section{Discussion}
\label{sct:discuss}
In this Section, we interpret the benchmarking results obtained with {\tt encube} on the Swinburne Discovery Wall. 
By considering survey sizes, data load times, visualisation configurations and interaction frame rates, we estimate the visualisation throughput, which we present in terms of the number of sources that could be examined in a given period of time. As a reflection on the role for bespoke visualisation environments in astronomy,  we also discuss the evolution of advanced visualisation systems when used in astronomical research projects.

\subsection{Load times}
\label{sct:load}
In order to be a useful adjunct to desktop-based visualisation methods, an alternative display solution needs to provide an appropriate level of computational performance.

Regardless of whether a single spectral cube or multiple cubes are to be visualised, there is an unavoidable overhead while the data is transferred from its storage location into the computer memory.  While this latency may not be as noticeable when working with a single cube, there is a cumulative loss of time when working with large surveys.  This effect increases if individual cubes are loaded multiple times for comparative tasks.  The most important factors in the load time are the network and internal transfer bandwidths and the volume of data.

Our benchmarking results revealed a strong positive correlation between $T_{\rm Load}$ and $V_{\rm Store}$ across a range of storage volumes from 1.17 GB to 34.73 GB.  This is consistent with our expectation that each of: (1) the data access and load phase, where each Process and Render node must transfer data via the NFS mount to the Master node; (2) the pre-computation performed for each spectral cube; and (3) the initial transfer of data to the GPU for texture-based volume rendering have $O(N)$ algorithmic behaviour. If any one of these processes imposed a bottleneck for the increasing total data volume, we would expect to see deviations away from the linear scaling. 

With the Swinburne Discovery Wall hardware, we can load 180 spectral cubes drawn from: (1) the LVHIS survey in under 2 minutes; (2) the WHISP survey in under 3 minutes; and (3) combinations of WHISP, THINGS and LVHIS cubes in under 5 minutes. 

Using the median $T_{\rm Load}$ for WHISP-only surveys in Table \ref{tbl:config}, we can consider alternative configurations that reach the same total number of data cubes, but through multiple loads of smaller quantities at a time.  An additional overhead here is that we need to wait $T_{\rm Socket} = 60$ seconds for the Process and Render nodes to release their socket connections before the next configuration can be loaded. Expected total load times (rounded up to the nearest half minute) are as follows:
\begin{itemize}
    \item Nine sets of 20 WHISP cubes will load in 11.5 minutes
    ($9 \times 21 + 8 * T_{\rm Socket} = 669$ s);
    \item Four sets of 40 WHISP cubes plus one set of 20 WHISP cubes will load in 7.0 minutes
    ($4 \times 38 + 1 \times 21 + 4 * T_{\rm Socket} = 413$ s); and
    \item Two sets of 80 WHISP cubes plus one set of 20 WHISP cubes will load in 5.0 minutes
    ($2 \times 73 + 1 \times 21 + 2 * T_{\rm Socket} = 287$ s).
\end{itemize}
By increasing the total number of cubes displayed on the Discovery Wall, we benefit from parallelisation across the Process and Render nodes during the pre-computation phase and we do not experience the system latency imposed by $T_{\rm socket}$.  The advantage of using the 4K UHD monitors is that we retain a reasonable image resolution per source even when there are 18 spectral cubes per individual monitor (36 cubes per column) of the Discovery Wall.

\subsection{Frame rates}
Once a configuration of spectral cubes has been loaded and displayed on the Discovery Wall, the most important metric is the frame rate.    The higher the frame rate, the smoother the interaction experience when modifying the location of the camera (e.g. when controlling the visualisation of all the spectral cubes simultaneously via the user interface).  

For {\tt encube}, there are several key observations that we make:
\begin{itemize}
    \item The frame rate depends on the size of the {\tt S2PLOT} window, such that expanding over both 4K-UHD monitors per Process and Render node decreases the frame rate.  This is seen in the per-column frame rates in Table \ref{tbl:framerate}, where $F_1$ values (the Master node) are generally higher than those of the other four columns ($F_2$ to $F_5$).  In order to display the user interface in the web browser on the Master node, we do not extend the {\tt S2PLOT} window across both monitors.
    \item There are variations in the frame rate as a function of viewing angle, which depends on the relative number of voxels along each axis of a cube [see, for comparison, Figure 5 of \citet{Hassan12}].  By reporting the lowest measured frame rates after each cube has undergone several complete rotations, we are presenting worst-case outcomes on interactivity.
    \item Frame rates can decrease when zooming in on details.  The amount of processing work performed by the GPU depends on the fraction of screen pixels that contain visible data.  When zoomed out, a larger percentage of each panel comprises non-data (i.e. background) pixels.  We did not record the effect on frame rates as the default configurations for 180 cubes presents a comparable ratio of data to total pixels as occurs when zooming in on with one of the lower $N_{\rm cube}$ configurations.
\end{itemize}
 
Setting a target of 10 frames/s as an indicator of reasonable interactivity with the data cubes, we exceed this for all of the 20-cube mock surveys (mean and median frame rates in Table \ref{tbl:framerate}), and for configurations of 180 sources selected entirely from the WHISP and LVHIS surveys.   

For the 180-cube combination configuration, which includes a randomly-selected sample of 60 THINGS cubes, the mean and median frame rates fall below 5 frames/s.  Here, the higher frame rates measured for spectral cubes assigned to the fifth column of the Discovery Wall (column $F_5$ in Table \ref{tbl:framerate}) occur as only 5-6 out of 36 spectral cubes were randomly selected from the THINGS survey. If we had ``perfect'' randomness in the construction of the mock survey samples, we would expect 12 THINGS galaxies assigned to each column.  Instead, columns two to four are required to perform much more processing than column five per screen refresh (more memory or total voxels per GPU), resulting in the lower frame rates for ($F_2$ -- $F_4$) when a single GPU is driving two 4K UHD monitors.

\subsection{Throughput}
\label{sct:through}
One of the key metrics we wish to ascertain is the visualisation throughput, $V_{\rm tp}$, which is the number of source cubelets that can be inspected in a given period of time, measured in units of sources/hour.

For a single user, it is not expected that a peak $V_{\rm tp}$ could be sustained throughout an entire day, but it is reasonable to assume that rates of $25-50\%$ of $V_{\rm tp}$ might be achievable for extended periods of time.  This is compatible with a work pattern for quality control or source-finding candidate rejection where the candidates from the latest large-scale survey cube(s) are assessed daily. 

\subsubsection{Multi-object workflows}
To estimate the throughput for a multi-object workflow, we consider two scenarios using the combination mock survey:
\begin{itemize}
    \item An 80-cube configuration. The full dataset loads in around $T_{\rm Load}$ = 160 seconds (mean load time plus one standard deviation). An initial inspection can occur in $T_{\rm Inspect}$ = 180 seconds (see Section \ref{sct:candidate}).  If we assume 25\% of sources require additional action, and the recording of that action takes 60 seconds, then $T_{\rm Action}$ =  1200 seconds.
    \item A 180-cube configuration. The full dataset loads in $T_{\rm Load}$ =  300 seconds.  The time required for the initial inspection is assumed to scale linearly with the number of sources, such that $T_{\rm Inspect} \sim$ 405 seconds. With 25\% of sources requiring a 60-second action to be recorded, then $T_{\rm Action}$ = 2700 seconds.
\end{itemize}

The total time required for the completion of a SIMD  process with {\tt encube} is then:
\begin{equation}
    T_{\rm SIMD} = T_{\rm Load} + T_{\rm Inspect} + T_{\rm Action} + T_{\rm Socket}
\end{equation}
where $T_{\rm Socket}$, introduced in Section \ref{sct:load}, is a system latency.  Using the values proposed for these four quantities, we suggest that $T_{\rm SIMD}(80 \, \mbox{cubes}) = 1600$ seconds (26.7 minutes) and $T_{\rm SIMD}(180 \, \mbox{cubes}) = 3465$ seconds (58 minutes).  

Taken together, we estimate that $V_{\rm tp}$ = 160-180 sources/hour seems reasonable for the completion of one of the three SIMD tasks we have considered in our VDAR study.  Moreover, we have assumed only a single astronomer completing the task, whereas the large-format workspace of the Discovery Wall comfortably accommodates a small group working together.

\begin{figure}[ht]
    \centering
    \includegraphics[width=8.5cm]{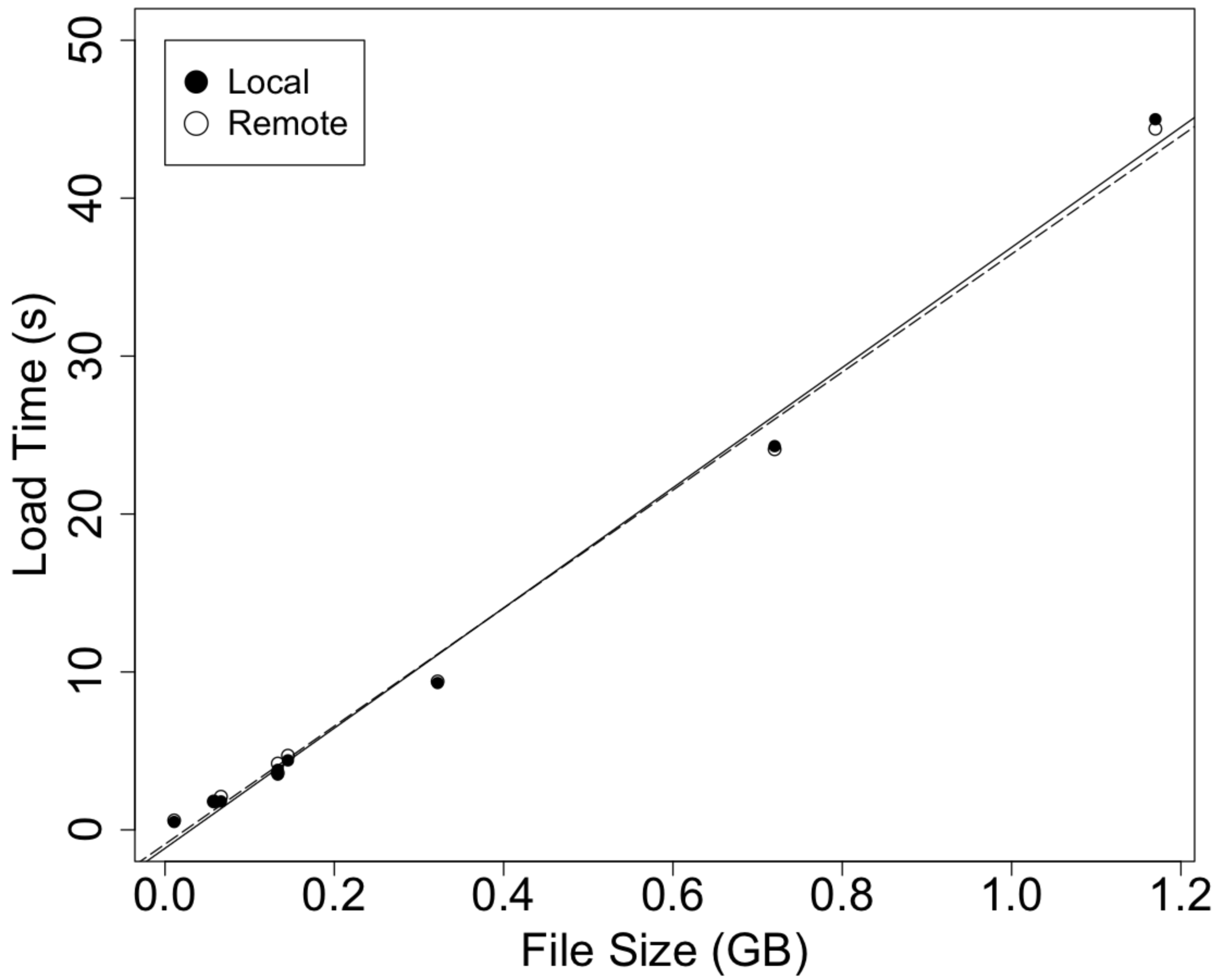}
    \caption{Single file load time for the three representative spectral data cubes (minimum, median and maximum file sizes) for each of the WHISP, THINGS and LVHIS surveys. Load times were measured for the local disk (filled circles) and across the local network via an NFS mount (open circles). In both cases, there is minimal difference between the two measurements, with a reaction time error of 0.5 seconds.}
    \label{fig:extra}
\end{figure}

\subsubsection{Comparison with single-object workflows}
As a point of comparison, we consider a single-object workflow, i.e. one source is loaded and visualised at a time with {\tt encube} and using the Swinburne Discovery Wall hardware.

A relationship between the single object load time and the FITS filesize was determined using a minimal sample of representative spectral cubes from each of the WHISP, THINGS and LVHIS datasets. We select the cubes with the smallest and largest filesizes, along with a cube that had the median file size (see Table \ref{tbl::surveys}). We measure load times for visualisation with {\tt encube} running only on the head node, where the data is stored, and on a remote machine over the network via the NFS mount. We used a manual timing method with a reaction time error of 0.5 seconds. 

As shown in Figure \ref{fig:extra}, we find minimal differences in load times from the local disk (filled circles) or via the remote NFS mount (open circles).  Performing a least squares fit to the combined data, we obtain:
\begin{equation}
    T_{\rm Load} = 37.71 V_{\rm Store} - 1.04 \; \rm{seconds}
    \label{eqn:singleload}
\end{equation}
with a Pearson correlation coefficient between $T_{\rm Load}$ and $V_{\rm Store}$ calculated to be $r = 0.997$.

\begin{table}[ht]
\caption{Single-object (Single) and multi-object (Multi) mean and median load times, $T_{\rm Load}$ in seconds, for the 80-cube [W]HISP, [T]HINGS, [L]VHIS and [C]ombination configuration, using survey data volumes from Table \ref{tbl:config}. The ratio of the single-to-multi object load times are recorded in the final two columns.}
\label{tbl:singlefile}
    \begin{tabular}{crrrrcc}
&    \multicolumn{2}{c}{\bf{Single} $\mathbf{T_{\rm Load}}$\bf{(s)}} & \multicolumn{2}{c}{\bf{Multi} $\mathbf{T_{\rm Load}}$\bf{(s)}}  & \multicolumn{2}{c}{\bf{Ratio}}  \\
Survey & Mean & Median & Mean & Median & Mean & Median \\
        \hline
W & 73 & 74 & 363 & 361 & 4.97 & 4.88 \\
T & 271 & 271 & 1201 & 1194 & 4.44 & 4.41 \\
L & 57 & 57 & 238 & 238 & 4.17 & 4.17 \\
C & 148 & 146 & 581 & 580 & 3.93 & 3.97
    \end{tabular}
\end{table}

\begin{figure}[ht]
        \centering
\includegraphics[width=8.5cm]{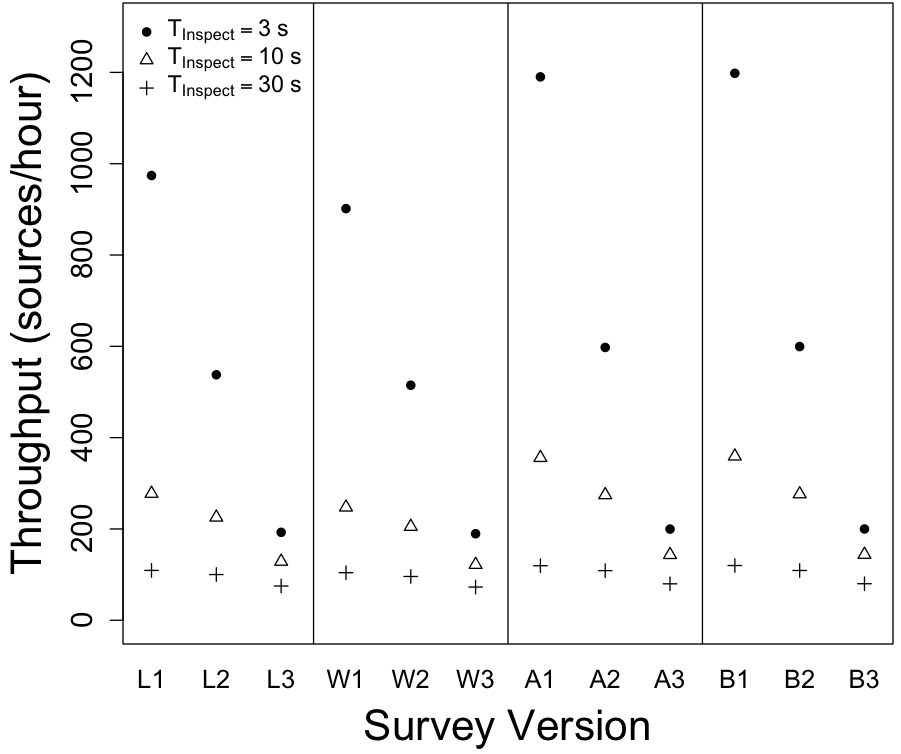}
    \caption{Estimated throughput for a SIMD workflow based on visual inspection of the entire [L]VHIS, [W]HISP, [A]PERTIF and WALLA[B]Y extragalactic H{\sc i} surveys, as per configurations described in Section \ref{sct:aperby}.For each survey, we consider three scenarios with different follow-up action times: (1) $T_{\rm Action} = 0$; 
 (2) $T_{\rm Action} = 30$ s/source for $10\%$ of sources; and (3) $T_{\rm Action} = 60$ s/source for $25\%$ of sources. Symbols are used to differentiate between the inspection times, with $T_{\rm Inspect} = 3$ s/source for a multi-object workflow (filled circle) and $T_{\rm Inspect} = 10$ s/source (open triangle) and $T_{\rm Inspect} = 30$ s/source (plus symbol) for single-object workflows.}
    \label{fig:samplesize}
\end{figure}

Using the average and median sample survey file sizes from Table \ref{tbl:config}, we compare the single-object and multi-object load times for the 80-cube WHISP, THINGS, LVHIS and combination configurations -- see Table \ref{tbl:singlefile}. The ratio of the single-to-multi object load times was calculated for each configuration, showing a $4-5$ times speed-up in load times using the five compute nodes of the Swinburne Discovery Wall.  This is not surprising for the nearly-perfect parallelism expected in this stage of the workflow, but with a slight input/output bottleneck at the head node where all of the data is stored.

\subsubsection{Estimates for future extragalactic H{\sc i} surveys}
\label{sct:aperby}
In Figure \ref{fig:samplesize}, we estimate and compare the throughput for multi-object and single-object SIMD workflows.  In addition to the LVHIS and WHISP extragalactic H{\sc i}, we obtain preliminary results for the APERTIF and WALLABY surveys; these values are indicative only of future analysis that is yet to be completed.  We base our throughput predictions on 10,000 APERTIF sources (in the velocity range 1,000 to 10,000 km/s) with a mean storage volume of 0.62 MB/source cubelet\footnote{K.Hess, private communication} and 210,000 sources in WALLABY with a mean storage volume of 3 MB/source cubelet.\footnote{Analysis by author CM}

The time to inspect each source is highly dependent on the SIMD task.  For the candidate rejection VDAR activity (Section \ref{sct:candidate}), we performed an initial visual scan across 80 spectral data cubes displayed on the Swinburne Discovery Wall in three minutes or 2.25 seconds/cube.  This is achievable once all cubes have been loaded using physical navigation to rapidly move around the display space.  With the continual cognitive set-shifting required for a lone astronomer to load and inspect one cube at a time, regardless of the display and visualisation software used, it may take 10-30 seconds per cube even at peak performance.  Moreover, the single-object workflow removes the opportunity to perform comparisons, or rapid revisits to double check that a previously-viewed source had been inspected adequately.

For each survey, we consider three scenarios with different follow-up action times: (1) $T_{\rm Action} = 0$, such that inspection occurs but no additional actions are required for all sources; 
 (2) $T_{\rm Action} = 30$ s/source for $10\%$ of sources; and (3) $T_{\rm Action} = 60$ s/source for $25\%$ of sources.  Symbols are used in Figure \ref{fig:samplesize} to differentiate between the inspection times, with $T_{\rm Inspect} = 3$ s/source for a multi-object workflow (filled circle) and $T_{\rm Inspect} = 10$ s/source (open triangle) and $T_{\rm Inspect} = 30$ s/source (plus symbol) for single-object workflows.  For large survey sizes, $N_S$, these components of $T_{\rm SIMD}$ dominate over $T_{\rm Load}$ regardless of whether a single-object or multi-object workflow is used. The minor contribution from $T_{\rm Socket}$ has been omitted. In all of the scenarios we considered, the estimated throughput with a multi-object workflow exceeds that of a single-object workflow. 
 
\subsection{Evolution of visualisation solutions}
\label{sct:evolution}
Astronomers have developed their craft over centuries by using a combination of singular, bespoke facilities for data gathering (e.g. dedicated observatories and supercomputers) supported by widely-available, general purpose resources for data analysis and visualisation (e.g. desktop and laptop computers in the digital era). We assert that a complementary role exists for dedicated advanced visualisation facilities that can provide a very different experience to that of the everyday.  

In the same way that astronomers do not expect to operate their own personal 64-metre radio telescope or 8-metre class optical/infrared telescope, there should not be an expectation, or need, for {\em all} astronomical institutions to operate a local advanced visualisation facility.  What is more important is that when such facilities are available, there is a community of interested and potential users who are able to take advantage of them.

As astronomical teams prepare themselves for the next phase of petascale and exascale data collection, new visualisation strategies that enable and enhance survey-scale discovery-based research processes will be required.   Our VDAR evaluation demonstrates how comparative visualisation (implemented using {\tt encube} and the Swinburne Discovery Wall) could be applied to SIMD visual analysis tasks that would not otherwise be feasible using a standard desktop configuration. 

Until a survey project is underway, the exact configuration of software and hardware that provides the most productive approach to advancing scientific knowledge may not be known.  As the projects develop, familiarity with the strengths and weaknesses of the instrumentation and software-pipelines will also grow.   The strategies for analysis and visualisation adopted during the first year of data collection may not be the same as those deemed essential in the years that follow.

Some approaches to analysis and visualisation become essential throughout the lifetime of the individual research project where they were first adopted, perhaps spreading further into the discipline to become ubiquitous. Other alternatives may be relevant for a short period of time, or may only need to be accessed by a few members of a research team, but provide a much-needed distinctive perspective that serves to accelerate discovery.  By presenting alternatives to current ways of working, astronomers can consider for themselves whether a combination of options will assist them at various stages of their research workflow.  

As an illustrative example of the evolution in the use of display environments, we look to the real-time, multi-wavelength Deeper Wider Faster (DWF) fast transient detection program \citep{Andreoni19}, where the Swinburne Discovery Wall -- used as a TDW without {\tt encube} -- has also played an important role. 

As an international collaboration, DWF operations rely on a core team of co-located human inspectors with access to suitable visualisation software and hardware to support their decision-making processes during high-intensity, real-time observing campaigns. Through identification of potential fast or short-lived transient events, the DWF team determines whether there is a need to trigger immediate follow-up observations (e.g. target of opportunity spectroscopic observations with one of the Keck Observatory telescopes). 

Informed by a user performance study that investigated potential roles for TDWs in supporting inspection of very high pixel-count images by individuals or small teams \citep{Meade14}, a TDW became a necessary component of the display ecology used in the DWF project. The TDW replaced an initial inefficient visualisation workflow (used during pilot observations in 2015), where the research team used laptop screens and desktop monitors to inspect each of the 60 CCD frames (4096 $\times $ 2048 pixels) per field imaged with the Dark Energy Camera [DECam; \citet{Diehl12, Flaugher12}].  

Over successive observing campaigns, as reported by \citet{Meade17}, the role and configuration of the TDW changed in response to user requirements and feedback.  
The visual inspection tasks performed by DWF team members were modified due to improvements in scientific understanding of the categories of fast transients that were being identified in real-time (and by extension those categories that could be analysed after the short-duration observing campaigns had concluded), along with enhancements to the automated pipelines \citep{Andreoni17,Goode22}.
In turn, improvements of the automated pipeline were directly informed by the knowledge the team acquired through using the TDW.

At the time of writing, while no longer essential in the DWF context, the Swinburne Discovery Wall continues to play a role during real-time DWF campaigns.  At critical stages of the development of DWF, however, the TDW was a solution that was ``fit for purpose'' and supported team-based visual discovery tasks that were not feasible to conduct with a standard desktop-bound approach.

\section{Conclusions}
\label{sct:conclude}
The expected growth in both the volume and velocity of data from future  astronomical surveys necessitates a move away from serial workflows. 
The comparative visualisation approach we have investigated here via benchmarking and a VDAR evaluation is not intended to replace existing alternatives, but provides a demonstration of a complementary workflow that addresses some existing -- and emerging -- challenges in the size and scale of astronomical surveys.

Within our case study context of extragalactic H{\sc i} surveys, we anticipate that both the short and longer term use of automated pipelines will retain a stage of visual inspection and classification.  We suggest that this can be achieved more successfully, and more rapidly, using a method that is not about inspecting one object at a time. 

As we have shown here,  the {\tt encube} framework operating on a tiled display wall presents a compelling alternative mode for SIMD activities. 
We have considered tasks that are highly repetitive, yet may need to be performed on all sources detected within a survey.  Examples here include quality control, candidate rejection, and morphological classification.  In all cases, as identified through our VDAR studies, {\tt encube}
encouraged a sensemaking process \citep{Pirolli05} with a foraging phase and a sensemaking loop.  The comparative nature of the display -- comfortably visualising 180 spectral cubes at a time, using the Swinburne Discovery Wall configuration of ten 4K-UHD monitors -- supports the rapid identification of features affecting multiple source cubelets while also presenting immediate access to both the spatial and spectral data for individual objects (through our use of volume rendering).

A few hours interacting with data with {\tt encube} on the Discovery Wall could replace weeks to months of work at the desktop -- without diminishing the importance of the follow-up detailed analysis that the desktop supports. We estimate a throughput of 160-180 sources/hour could be inspected using the configuration that we assessed.  

Both {\tt encube} and the Swinburne Discovery Wall are easily modifiable and scalable, in the sense that additional columns of monitors plus computers can be added to increase the number of sources displayed at a time. Implementation of our solution at another institution requires access to: the open-source software\citep{encube2017ascl.soft06007V}; one or more Linux-based computers; (ideally) multiple monitors; and an appropriate network connection between the process and render nodes and the master node where the data set is stored.

Customised visualisation and analysis approaches will evolve over time as  surveys progress.  They should be employed during those periods that are particularly labour-intensive, while assisting in the identification of additional processes that can be fully or partly automated.  Finding the appropriate balance between human inspection and automated detection may help to maximise the overall discovery potential of a workflow \citep{Fluke17,Fluke2020}.

\section*{Acknowledgements}
We acknowledge the Wurundjeri People of the Kulin Nation, who are the Traditional Owners of the land on which the research activities were undertaken.
Christopher Fluke is the SmartSat Cooperative Research Centre (CRC) Professorial Chair of space system real-time data fusion, integration and cognition.  SmartSat CRC's activities are funded by the Australian Government's CRC Program.  We acknowledge the generous support of the Eric Ormond Baker Charitable fund, which helped to establish the Discovery Wall and the remote observing facility at Swinburne University of Technology. We are extremely grateful to David Barnes and Amr Hassan for their technical advice and encouragement during early phases of this work, and to Kelley Hess for assisting with understanding the preliminary APERTIF H{\sc i} survey results.  This paper made use of data from: WHISP, Westerbork Observations of Neutral Hydrogen in Irregular and Spiral Galaxies \citep{Hulst2001ASPC..240..451V,Swaters02}; THINGS, The H{\sc i} Nearby Galaxy Survey \citep{Walter2008AJ}; and LVHIS, The Local Volume H{\sc i} Survey \citep{Koribalski2018MNRAS.478.1611K}.

\bibliography{references} 

\begin{appendix}

\begin{figure}[ht]
    \centering
    \includegraphics[width=8.5cm]{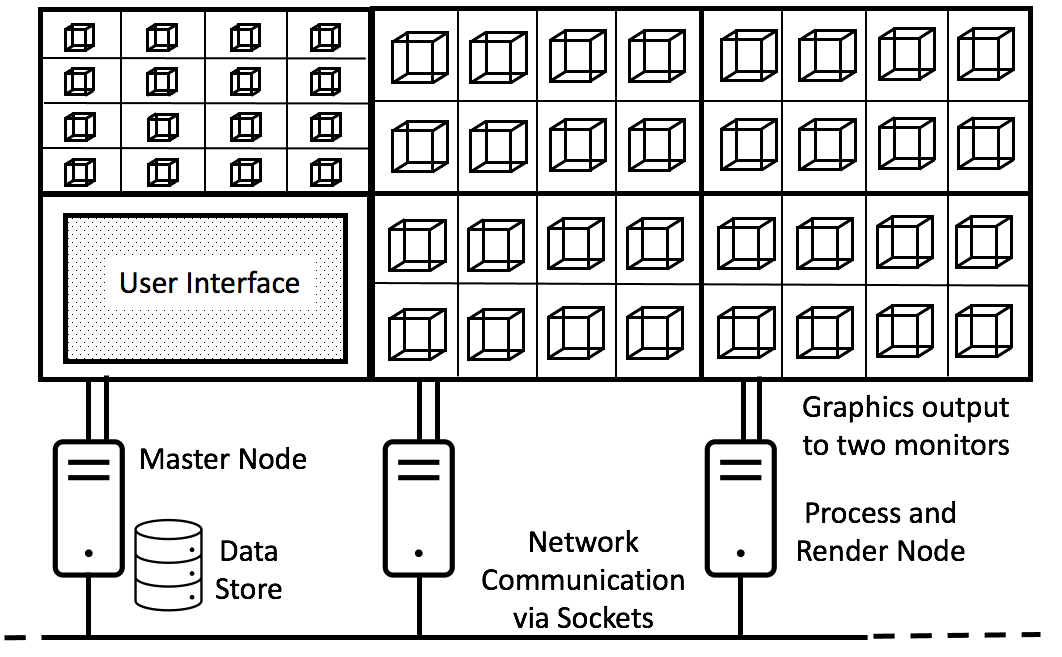}
    \caption{The key components required for {\tt encube} to operate on the Swinburne Discovery Wall.  The Master node hosts the Data Store, which is accessed by the Process and Render nodes via a network file system mount point.  Direct communication between the Process and Render nodes and the Master occur over the shared network via sockets.  Each Process and Render node provides a graphical output to two monitors, which are tiled into a matrix of {\tt S2PLOT} panels.  The User Interface operates on the Master node, controlling the assignment of spectral cubes to each of the Process and Render nodes and modification of the appearance of the spectral cubes. }
    \label{fig:layout}
\end{figure}

\begin{figure}[ht]
\centering
   \includegraphics[width=8.5cm]{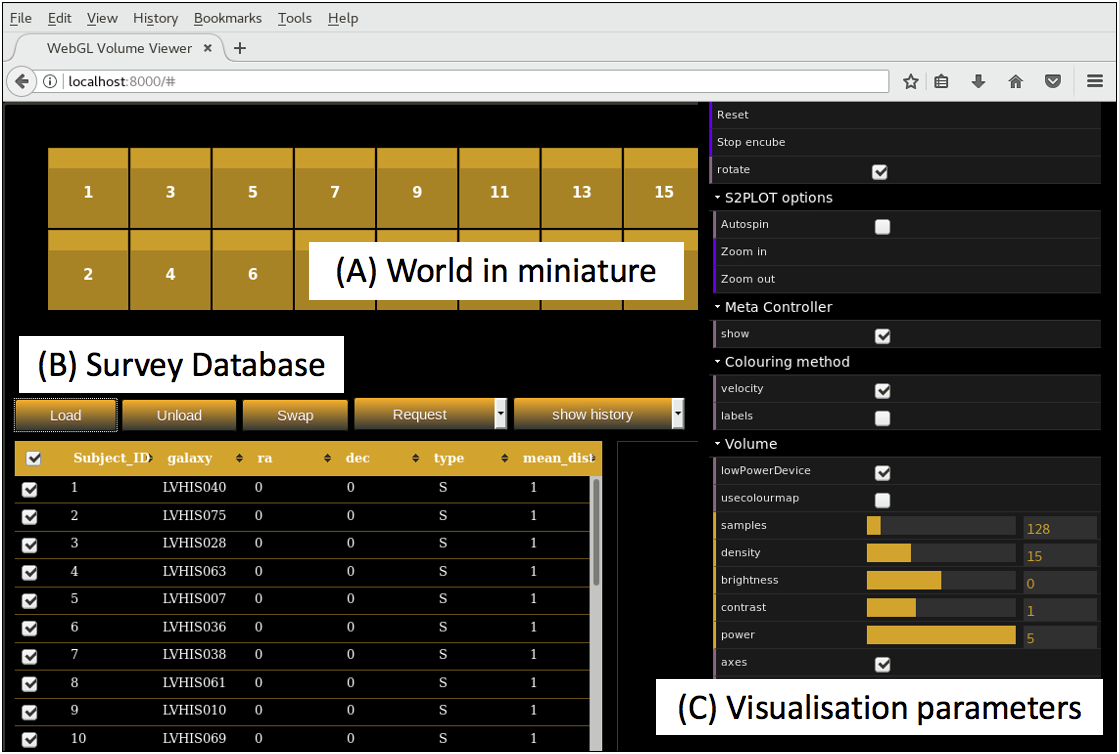}
    \caption{The {\tt encube} user interface (UI) operating in the Firefox Web browser on the Master node.  The main elements of the UI are: (A) the world in miniature view, replicating the layout of the Discovery Wall; (B) the survey database containing filenames and associated metadata; and (C) the visualisation parameters, controlling visual aspects such as choice of colour map and labelling of spectral cubes.  Additional section of the interface (not shown here) includes the camera controller and interactive plots such as voxels histogram (i.e. to modified the dynamic range) or other custom meta information (e.g. stellar masses of galaxies displayed on the screens as a function of grid position).}
    \label{fig:interface}
\end{figure}

\section{Implementation notes}
\label{app:technical}
\subsection{Technical matters}
\label{sub:keycomp}
In this section, we highlight some additional features of the implementation of {\tt encube} on the Swinburne Discovery Wall.   One workstation is assigned the role of the Master Node, where the manager unit and interaction unit  are deployed.  All five workstations act as Process and Render nodes.  Figure \ref{fig:layout} illustrates the connections and communication pathways between the Master node and each of the Process and Render nodes.

\begin{figure*}
    \centering
    \includegraphics[width=18cm]{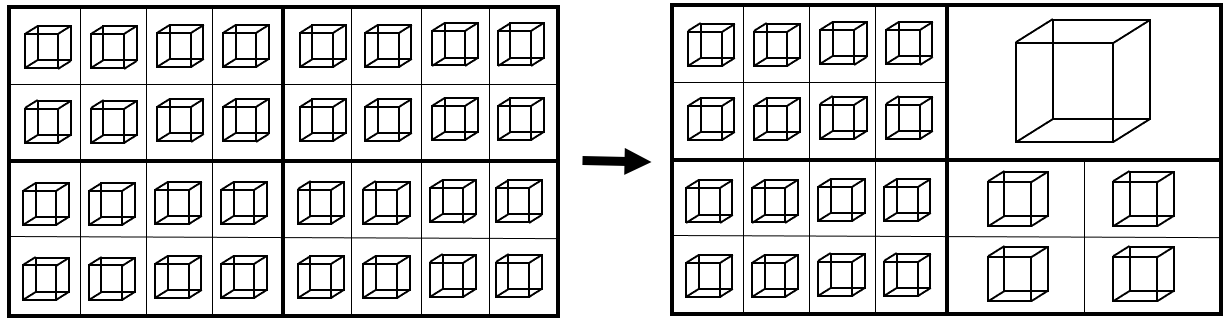}
    \caption{A proposed enhancement to {\tt encube} would support non-uniform tiling of the display area.   In the existing configuration (left-hand panel), the same level of detail is used for every spectral cube.  A modification to the tiling (right-hand panel) would allow individual cubes with different sizes to be presented at the same scale or for the volume rendering to occur with a higher level of detail.}
    \label{fig:tiling}
\end{figure*}

{\tt Encube} is launched from a Linux terminal on the Master node, which activates the program instance on each of the Process and Render nodes.   Each program instance: (1) creates and opens a socket for communication with the Master node; (2) and makes application programming interface (API) calls in C code to the {\tt S2PLOT} library for interactive graphical elements.   Relevant content from the configuration file hosted on the Master node is passed to the Process and Render nodes.  Once the socket connections have been established, the user interface is accessed through a Web browser accessing {\tt localhost} on the Master node (see Figure \ref{fig:interface}).

{\tt S2PLOT} allows for the creation of independent regions of the graphics display window, referred to as panels.   For simplicity, panels are presented in {\tt encube} as a uniformly tiled matrix of rows and columns.   The 3D geometry within an {\tt S2PLOT} panel can be controlled by selecting the panel and using the attached mouse to rotate the data cube or the keyboard to zoom in or out.  As each display column of the Discovery Wall is independent, it is possible to use the keyboard and mouse associated with a column in order to work with a local subset of data (see Figure \ref{fig:sdw}).  Alternatively, the location, orientation and view direction of the virtual camera can be set for each panel using an API call. This method is used when interacting with the user interface on the Master node, so that the virtual camera is updated simultaneously for all of the panels.    

Each Process and Render node requests and loads relevant data files from the Master node, using a drive that is accessible using the network file system (NFS).  Once each Process and Render node has loaded the required data, the spectral cube is visualised using 3D texture-based volume rendering.  Here, an {\tt S2PLOT} callback function is associated with each panel, and once per refresh cycle, the volume rendering is generated based on the current virtual camera position.   
3D texture-based rendering provides a compromise between lower-fidelity two-dimensional texture image stacks (also implemented in {\tt S2PLOT}) or computationally-demanding ray-shooting.  

For simplicity of operation, two different colour-mapping options are provided: intensity-based, whereby a heat-style colour map is assigned from the minimum to the maximum voxel value for each spectral cube, and velocity-based mapping \citep{vohl2017MNRAS.471.3323V}.  Here, the velocity data is utilised along with the voxel values, in order to provide cues as to whether neutral H{\sc i} gas is blue-shifted or red-shifted along the spectral axis with respect to the centre of the cube (assumed to be equivalent to the centre-of-mass for most systems).

While completing the benchmarking and VDAR evalauation activities (described in Sections \ref{sct:benchmarks} and \ref{sct:vdar}), we chose not to invest development time to make some cosmetic changes to the {\tt encube} user interface.   In particular, the world in miniature component of the interface (see Figure \ref{fig:interface}) was not ideal when the number of spectral cubes visualised exceeded 40. This temporarily limits the ability to use some of the features of {\tt encube}, such as the ability to select and swap cubes between any of the displays in real-time.  However, the overall functionality and performance of the {\tt encube} process and render components is not impeded. 

In the implementation of {\tt encube} that we benchmarked, there were some additional processing steps performed that add to the time taken to load each spectral cube. These comprise several independent complete passes through the spectral cube to calculate statistical parameters, compare actual data values with those recorded in the spectral cube metadata, and generation of a histogram of data values for each spectral cube.  Each of these processes have algorithmic linear scaling depending only on the number of voxels in the spectral cube. Consequently, they introduce a multiplicative factor on the time to load all of the spectral cubes. Such pre-computation is a design choice that allows the CPU memory to be freed once data is loaded onto a GPU. Accessing these values has $O(1)$ complexity later during interactive analysis.  

\subsection{Future enhancements}
While working with {\tt encube} during the VDAR evaluation, we identified several additional features or enhancements that could extend the framework's suitability for comparative visual analysis of large-scale extragalactic H{\sc i} surveys:
\begin{itemize}
\item Add an on-screen scale indicator. As all spectral cubes are scaled to a unit cube for convenience, the physical size of individual objects was lost.
\item Within the user interface, allow selection or sorting of the source list by any metadata attribute, such as size, total H{\sc i} mass, or distance.
\item Access and display detailed metadata of a selected object or set of objects. During the present work, a trivial modification was made to toggle visibility of the name of each object within its {\tt S2PLOT} display panel.
\item Improve the creation of the on-screen configuration, allowing more flexibility in how data is assigned to the available display space. For example, a non-uniform arrangement of panels per column, which could allow individual spectral cubes to be visualised at increased levels of detail or cubes with different sizes (e.g. spatial pixel coverage or rest-frame physical dimensions) could be  presented at the same scale as demonstrated in Figure \ref{fig:tiling}. 
\item Include support for additional data types to be loaded and displayed, including spectral cubes from different wavelength regimes or observing modes (e.g. optical integral field units), overlay of two-dimensional images, or visualisation of one-dimensional spectra.
\item Provide a mechanism by which annotations could be recorded regarding individual sources, preferably through the use of speech-to-text capture and conversion.
\item Support interactive masking of channels via the user interface for selected subsets of cubelets, so that the issues identified with the WHISP sample could have been resolved in real-time.  Such modifications could then be embedded into the dataset, by exporting the modified spectral cubes for future automated, or human, analysis.
\end{itemize}

\end{appendix}

\label{lastpage}
\end{document}